\documentclass[journal]{vgtc}                
\usepackage[dvipsnames]{xcolor}
\usepackage{amsmath,amsfonts,amssymb}
\usepackage[english]{babel}    
\usepackage{url}
\usepackage{chngcntr}          
\usepackage{listings}
\usepackage{comment}
\usepackage{gensymb}  
\usepackage{booktabs} 
\usepackage{siunitx} 
\usepackage{changepage}

\usepackage{enumitem}
\usepackage[
singlelinecheck=false 
]{caption} 

\ieeedoi{10.1109/TVCG.2019.2934536}

\addto{\captionsenglish}{}

\addto\captionsenglish{}  

\DeclareMathAlphabet{\mathsfit}{T1}{\sfdefault}{\mddefault}{\sldefault} 

\ifpdf
  \pdfoutput=1\relax                   
  \usepackage{graphicx}                
  \DeclareGraphicsExtensions{.pdf,.png,.jpg,.jpeg} 
\else
  \usepackage{graphicx}                
  \DeclareGraphicsExtensions{.eps}     
\fi%

\graphicspath{{figures/}{pictures/}{images/}{./}} 

\usepackage{microtype}                 
\PassOptionsToPackage{warn}{textcomp}  
\usepackage{textcomp}                  
\usepackage{mathptmx}                  
\usepackage{times}                     
\usepackage{cite}                      
\usepackage{tabu}                      
\usepackage{booktabs}                  

\onlineid{1119}

\vgtccategory{Research}
\vgtcpapertype{Evaluation}

\title{
Unifying Effects of Direct and Relational Associations for Visual Communication}

\author{ Melissa A. Schoenlein, Johnny Campos, Kevin J. Lande, Laurent Lessard, and Karen B. Schloss}

\authorfooter{

\item  Melissa A. Schoenlein, Psychology and Wisconsin Institute for Discovery, University of Wisconsin--Madison. Email: schoenlein@wisc.edu.
\item Johnny Campos, Cognitive Science, University of California, Merced. Email: jcampos54@ucmerced.edu.
\item Kevin J. Lande, Philosophy and Centre for Vision Research, York University. Email: lande@yorku.ca
\item Laurent Lessard, Mechanical and Industrial Engineering, Northeastern University. Email: l.lessard@northeastern.edu
\item  Karen B. Schloss, Psychology and Wisconsin Institute for Discovery, University of Wisconsin--Madison. Email: kschloss@wisc.edu.
}

\shortauthortitle{Schoenlein \MakeLowercase{\textit{et~al.}}: Unifying Effects of Direct and Relational Associations for Visual Communication}

\abstract{People have expectations about how colors map to concepts in visualizations, and they are better at interpreting visualizations that match their expectations. Traditionally, studies on these expectations  (\textit{inferred mappings}) distinguished distinct factors relevant for visualizations of categorical vs. continuous information. Studies on categorical information focused on direct associations (e.g., mangos are associated with yellows) whereas studies on continuous information focused on relational associations (e.g., darker colors map to larger quantities; dark-is-more bias). We unite these two areas within a single framework of assignment inference. Assignment inference is the process by which people infer mappings between perceptual features and concepts represented in encoding systems. Observers infer globally optimal assignments by maximizing the ``merit,'' or ``goodness,'' of each possible assignment. Previous work on assignment inference focused on visualizations of categorical information. We extend this approach to visualizations of continuous data by (a) broadening the notion of merit to include relational associations and (b) developing a method for combining multiple (sometimes conflicting) sources of merit to predict people’s inferred mappings. We developed and tested our model on data from experiments in which participants interpreted colormap data visualizations, representing fictitious data about environmental concepts (sunshine, shade, wild fire, ocean water, glacial ice). We found both direct and relational associations contribute independently to inferred mappings. These results can be used to optimize visualization design to facilitate visual communication.
}

\keywords{Visual reasoning, information visualization, colormap data visualizations, visual encoding, color cognition}

 \teaser{
   \centering
   \includegraphics[width=1.0\linewidth]{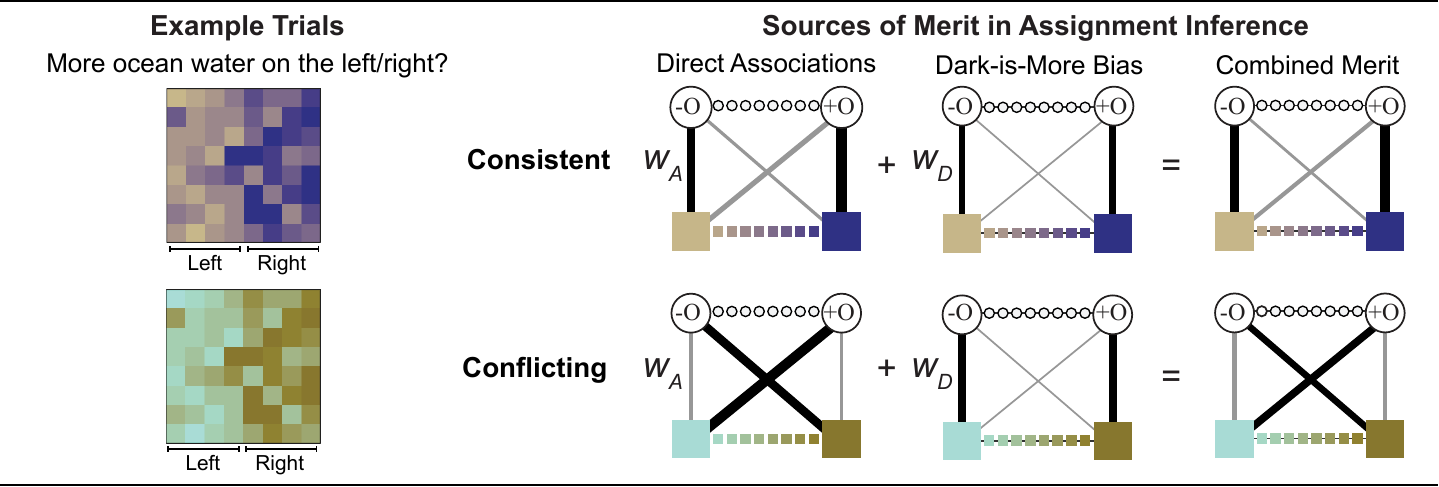}
   \caption{In this study, participants inferred which region of colormaps (left/right) represented more of a domain concept (e.g., ocean water). Inferences can be predicted by simulating assignment inference using a weighted combination of multiple (sometimes competing) sources of ``merit'': direct associations and relational associations (dark-is-more bias).}
   \label{fig:colormaps}
 }

\begin{document}


\firstsection{Introduction} \label{sec:intro}
\maketitle
Imagine you are interpreting a bar chart and need to infer which colors map to which concepts represented in the chart. Now, imagine instead interpreting a colormap data visualization\footnote{Literature on visualizing continuous data using color has inconsistent terminology. In this paper, ``colormap'' refers to a visualization that maps gradations of colors to quantities (e.g., weather maps, neuroimaging maps, correlation matrices). ``Color scale'' refers to the color gradient used to construct a colormap.} and you need to infer which colors map to which quantities represented in the colormap.

Traditionally, researchers studying the role of color semantics for visual communication have treated these cases as two distinct problems. One involves mapping colors to different categories in categorical information \cite{lin2013, setlur2016, schloss2018, schloss2021, mukherjee2022} and the other involves mapping gradations of color to gradations of quantity in continuous data \cite{cuff1973, mcgranaghan1989, schloss2019,  sibrel2020}. In both cases, a key goal is to understand people's expectations about the mappings between colors and concepts in visualizations (called \textbf{\textit{inferred mappings}}) because visualizations designed to match people's expectations are easier to interpret \cite{norman2013, tversky2002, tversky2011, hegarty2011, lin2013, schloss2018, schloss2019, schloss2021, sibrel2020, mukherjee2022}.

Studies on visualizations of categorical information focus on \textit{\textbf{direct associations}}---the degree to which each color is associated with each concept represented in the visualization. Methods have been developed to use direct associations to optimize mappings between discrete colors and concepts to facilitate visualization interpretability  \cite{lin2013, setlur2016, schloss2018, schloss2021, mukherjee2022}.

Studies on visualizations of continuous data focus on  \textit{\textbf{relational associations}}---correspondences between relational properties of visual features and relational properties of concepts. For example, observers have a dark-is-more bias, inferring that darker colors map to larger quantities \cite{mcgranaghan1989, cuff1973, brewer1994, schloss2019, sibrel2020}. This bias is relational because it depends on relative lightness, rather than particular colors in visualizations. Although empirical studies of colormaps have focused on relational associations and explicitly tried to avoid potential effects of direct associations \cite{mcgranaghan1989, schloss2019, sibrel2020}, direct associations likely play an important role (see Samsel et al.'s \cite{samsel2017} intuitive colormaps for environmental visualizations). 

In this paper, we aim to unite the study of direct and relational associations under a single framework of \textit{\textbf{assignment inference}}. Assignment inference is the process by which people infer mappings among visual features and concepts in visual encoding systems \cite{schloss2018}. Previous work on assignment inference focused on visualizations of categorical information, showing that observers infer optimal assignments (i.e., mappings) that maximize the total ``goodness'' of each possible color-concept pair \cite{schloss2018, schloss2021, mukherjee2022}. This ``goodness'' is called \textbf{\textit{merit}}.

We propose that assignment inference also governs inferences about the meanings of colors in visualizations of continuous data. In testing this possibility, our work makes the following contributions: (1) We broaden the notion of ``merit'' in assignment inference to include relational associations, and show that both relational and direct associations influence inferred mappings for colormap visualizations. (2) We develop a method for combining multiple (sometimes conflicting) sources of merit for simulating assignment inference, and show that our method effectively predicts inferred mappings for colormap visualizations. 

\section{Background} \label{sec:background}
In this section, we review previous work on color semantics in information visualization. Following tradition, we discuss effects of direct associations for  visualizations of categorical information and relational associations for visualizations of continuous data. We will unite these two areas in Section 3 on our approach in the present study.

\subsection{Direct associations and assignment inference}\label{sec:directAssign}
Direct associations (a.k.a. color-concept associations) are the degree to which a color is associated with a concept. They are estimated using various measures, including human judgments \cite{adams1973, ou2004, schloss2020blue,
jonauskaite2019machine, tham2020systematic, rathore2020,  schloss2021, schloss2018, murthy2022}, image statistics \cite{lin2013, setlur2016, lindner2012a, rathore2020}, and language corpora \cite{setlur2016, havasi2010}.  

\begin{figure}[htb]
 \centering
 \includegraphics[width=1.0 \linewidth]{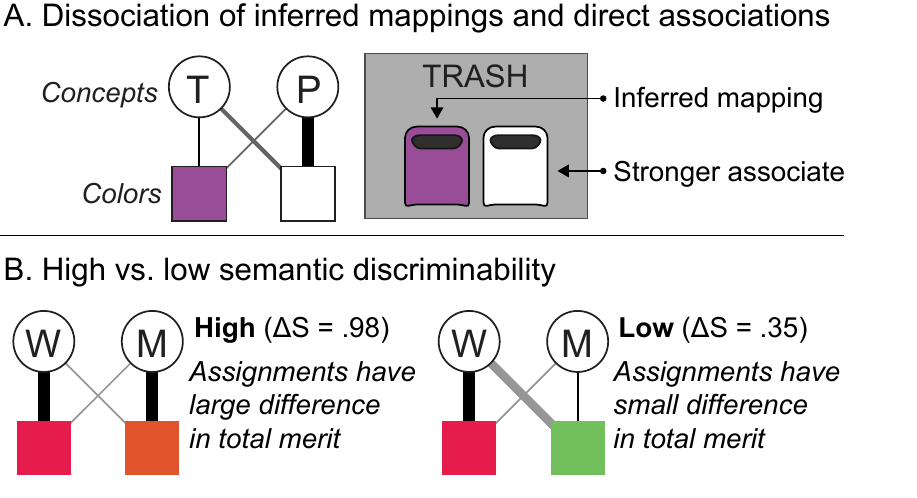}
 \vspace{-6mm}
 \caption{(A) Example of dissociation between associations and inferred mappings from \cite{schloss2018} (figure adapted from \cite{mukherjee2022}). Bipartite graph represents associations for concepts trash (T) and paper (P) with colors purple and white (thicker edges mean stronger associations). Observers infer trash maps to purple even though trash is more strongly associated with white, which is the optimal global assignment. (B) Color pairs with high vs. low semantic discriminability for concepts watermelon (W) and mango (M) from \cite{schloss2021}. $\Delta S$ indicates semantic distance. }
 \label{fig:assignInfBipartites}
 \vspace{-6mm}
\end{figure}

Although direct associations influence inferred mappings between colors and concepts in visualizations of categorical information \cite{lin2013, schloss2018, schloss2021, mukherjee2022}, direct associations and inferred mappings are not the same thing. Cases arise in which people infer that a concept maps to a weakly associated color, even when there are more strongly associated colors in a visualization. This distinction is shown in Fig. \ref{fig:assignInfBipartites}A. The bipartite graph (left) represents association strengths between each of two colors (purple and white) and each of two concepts (trash (T) and paper (P)) in an encoding system for recycling bins\cite{schloss2018}. The thickness of the edges connecting colors and concepts represents direct association strength (thicker indicates stronger association). Trash is more associated with white than with purple (thicker edges). Yet, when asked which colored bin is for trash (Fig. \ref{fig:assignInfBipartites}A right), people choose purple. Why?

Evidence suggests the reason is that people approach this problem using assignment inference, a process that considers all colors and concepts in the scope of the encoding system \cite{schloss2018}. Assignment inference is analogous to solving an assignment problem in optimization \cite{munkres1957algorithms}. In Fig. \ref{fig:assignInfBipartites}A, the scope of the encoding system includes trash and paper, even though paper was not relevant on this particular trial. Assignment inference does not simply assign a color to the concept with the strongest merit (for now, think of merit as direct association strength).  Instead, the process selects the combination of color-concept pairs that maximizes \textit{total} merit across all pairings.  The total merit for the T-purple/P-white assignment is greater than the alternative, T-white/P-purple. Thus, observers infer that trash maps to purple, despite trash being more strongly associated with white.

The ability to perform assignment inference depends on \textbf{\textit{semantic discriminability}} of the colors, given the concepts in the encoding system. Semantic discriminability can be understood by analogy to perceptual discriminability. Perceptual discriminability concerns how well one can distinguish the appearance of different colors, whereas semantic discriminability concerns how well one can distinguish the meaning of different colors in the context of an encoding system \cite{schloss2021, mukherjee2022}. In assignment inference, semantic discriminability is the degree to which one assignment has greater merit than the alternative assignment(s). For example, Fig. \ref{fig:assignInfBipartites}B shows color sets that differ in semantic discriminability for the concepts mango (M) and watermelon (W) (data from \cite{schloss2021}). The red and orange set (left) has high semantic discriminability because the W-red/M-orange assignment has far greater merit than the alternative. In contrast, the red and green set (right) has low semantic discriminability because the W-red/M-green assignment is only slightly better than the alternative. In their semantic discriminability theory, Mukherjee et al. \cite{mukherjee2022} specified constraints on the ability to design semantically discriminable color palettes for a given set of concepts. 

Semantic discriminability can be operationalized through a metric called \textit{\textbf{semantic distance ($\Delta S$)}} \cite{schloss2021, mukherjee2022}, which uses merit to quantify the degree to which any one assignment is better than alternative assignment(s), while accounting for uncertainty in the system. We reproduce the details for calculating semantic distance defined in \cite{schloss2021} in Supplementary Material Section \ref{sec:SuppSemDist} of the present paper. 

\textbf{Simulating assignment inference.} To simulate the outcome of assignment inference, it is necessary to (a) determine which assignment is optimal according to an assignment problem \cite{munkres1957algorithms} and (b) estimate the probability of inferring any one assignment over all alternative assignment(s), which is given by semantic distance. The combination of these two pieces of information indicates which colors observers will map to which concepts in assignment inference, and the probability that they will infer that assignment. This method is effective for predicting how people map colors to concepts for visualizations of categorical information (e.g., recycling bin signage \cite{schloss2018}, bar charts \cite{schloss2021, mukherjee2022}), although earlier work did not yet refer to ``semantic distance'' by name \cite{schloss2018}. This approach may also extend to inferences about properties of food and beverage products based on coloring in package design \cite{spence2022}.

\textbf{Definitions of merit for direct associations.} So far, we have treated merit merely as direct association strength. However, there are multiple methods to specify merit for direct associations, with some more effective than others \cite{schloss2018}. These different methods reduce to the same outcome in encoding systems with two concepts and two colors, such as those modeled in the present paper. Thus, we will withhold further discussion of metrics for computing merit for direct associations here, and we refer the interested reader to \cite{schloss2018} and \cite{mukherjee2022}.

\subsection{Relational associations}\label{sec:darkbiasB}
Relational associations are correspondences between relational properties of visual features (e.g., darkness, opacity, spatial arrangement) and relational properties of concepts (e.g., concepts of \textit{greater} or \textit{lesser} quantities). 
A fundamental aspect of relational associations is that they are structure-preserving. Structure preservation arises when structural properties between visual features correspond to structural properties among the concepts to which they are mapped \cite{palmer1978,shepard1970,blachowicz1997,maley2011,hegarty2011, goodwin2005, tversky2011}.

If particular relations among visual features are salient and certain relations among represented features are salient, then correspondences between these relations can be exploited to constrain the number of potential inferred mappings. For example, people are sensitive to the natural progression from lighter to darker shades and to the natural progression from smaller to larger quantities. Lightness can be mapped to quantities in many ways (see Fig. \ref{fig:structure_pres}A for four of many possibilities), but only two ways are structure preserving: darker colors map to larger quantities (dark-more assignment) or lighter colors map to larger quantities (light-more assignment).  From the perspective of structure preservation, both assignments are equally ``good.'' Any assignment that scrambles the mapping of lightness values to quantities is not structure-preserving and thereby is less ``good.'' 

Yet, not all structure-preserving assignments are equally good in peoples' inferred mappings. People have biases prioritizing one structure-preserving assignment over another \cite{mcgranaghan1989, cuff1973, schloss2019, sibrel2020}, discussed below. 

\textbf{Dark-is-more bias.}
The dark-is-more bias is the expectation that darker colors map to larger quantities (``more'' of what is being measured) \cite{mcgranaghan1989, cuff1973, schloss2019, sibrel2020}. People have a robust dark-is-more bias when interpreting colormaps without legends \cite{cuff1973, mcgranaghan1989}{\footnote{Although legends are a central part of colormap visualization grammar, Christen et al. \cite{christen2013} found that journal articles often leave out legends. Thus, studying colormaps without legends is relevant for real-world visualizations, while also providing a direct window into people's inferred mappings.}} and with legends \cite{schloss2019, sibrel2020}. Studying visualizations without legends, McGranaghan \cite{mcgranaghan1989} asked participants to interpret maps of U.S. states colored in shades of blue, and found that participants inferred that darker blues mapped to ``more.'' McGranaghan \cite{mcgranaghan1989} was purposefully ambiguous about the concept represented in the visualization, stating that the maps represented different amounts of ``data'' to avoid effects of direct color-concept associations. Studying visualizations with legends, Schloss et al. \cite{schloss2019}  presented participants with colormaps representing alien animal sightings, with the assumption that people would not have direct associations with these novel concepts. The legend either indicated dark-more encoding (greater animal sightings mapped to darker colors) or light-more encoding (greater animal sightings mapped to lighter colors). Overall, participants were faster at correctly interpreting the visualizations when legends indicated dark-more encoding, compared to light-more encoding, providing further evidence for the dark-is-more bias.

\textbf{Opaque-is-more bias.}
The opaque-is-more bias is the expectation that regions appearing more opaque represent larger quantities. This bias is only applicable when visualizations appear to vary in opacity \cite{schloss2019, bartel2021}, such as in value-by-alpha maps \cite{roth2010}. When the opaque-is-more bias is activated, it aligns with the dark-is-more bias on light backgrounds but conflicts with the dark-is-more bias on dark backgrounds. Under such conflicts, the opaque-is-more bias can cancel or even override the dark-is-more bias, leading observers to infer that lighter colors map to larger quantities \cite{schloss2019, bartel2021}. When the opaque-is-more bias is non-applicable (i.e., a visualization does not appear to vary in opacity), the dark-is-more bias leads observers to infer that darker colors map to larger quantities on both dark and light backgrounds \cite{schloss2019, bartel2021}. 

\textbf{Hotspot-is-more bias.}
The hotspot-is-more bias is the expectation that spatial regions that look like hotspots represent larger quantities in data. Hotspots emerge in datasets like fMRI, EEG, and meteorological data, in which extreme values are neighbored by less extreme values in concentric ring-like patterns \cite{schott2010}. Sibrel et al. \cite{sibrel2020} found that the dark-is-more bias dominated over the hotspot-is-more bias unless the hotspot was highly salient. Still, when colormaps contained hotspots that encoded larger quantities, they were easier to interpret when the hotspot was dark than when it was light (i.e., dark-is-more bias)\cite{sibrel2020}.

\section{Current approach} \label{sec:Approach}
Previous work on assignment inference focused on visualizations of categorical information, where merit depends on direct associations \cite{schloss2018, mukherjee2022, schloss2021}. We propose that assignment inference also governs inferred mappings for visualizations of continuous data, where merit may depend on both direct and relational associations. As such, assignment inference would operate over multiple (sometimes competing) sources of merit to determine inferred mappings. 

To test this possibility, we asked participants to infer the meanings of colors in colormaps (Fig. \ref{fig:colormaps}), and then predicted their responses using simulations of assignment inference. We studied inferred mappings for colormaps without legends, similar to \cite{cuff1973, mcgranaghan1989}.\footnote{This method requires fewer within-subject trials per condition than methods assessing inferred mappings for colormaps with legends, which require counterbalancing legend conditions (see \cite{schloss2019, sibrel2020}).} We assessed the proportion of times participants inferred the darker region mapped to ``more,'' depending on the domain concept and the color scales used to construct the colormap.  In Fig. \ref{fig:colormaps}, the domain concept is ocean water, and participants indicated whether there was more ocean water on the left or right of the maps. Colormaps were displayed on a white background and avoided hotspot spatial structure to prevent cases in which the dark-is-more bias conflicted with the opaque-is-more bias \cite{schloss2019} and hotspot-is-more bias \cite{sibrel2020}. In the General Discussion, we discuss extending our approach to handle these additional biases.

Next, we consider how direct and relational associations can serve as sources of merit for visualizations of continuous data, and how multiple sources combine to produce inferred mappings in assignment inference. 

\subsection{Direct associations as a source of merit}
Representing merit for direct associations in assignment inference for visualizations of continuous data (Fig. \ref{fig:colormaps}) is analogous to representing merit for direct associations for visualizations of categorical information (Fig. \ref{fig:assignInfBipartites}). In the examples in Fig. \ref{fig:colormaps}, merit from direct associations for the colormaps is illustrated in the bipartite graphs under the label ``direct associations.'' In the bipartite graphs, circles represent the endpoint concepts (more ocean water; +O, and less ocean water; -O) and squares represent the endpoint colors of color scales used to create the colormaps. Edge thickness represents association strength between each endpoint color and concept. From the perspective of merit from direct associations alone, assignment inference simulations for the colormaps in Fig. \ref{fig:colormaps} predict that more ocean water should map to darker blues in the top row and should map to lighter blues in the bottom row. 

Although the colormaps represent continuous data (more vs. less ocean water) with a continuous gradation of color, we simplify the assignment problem by focusing on only the the endpoint concepts and endpoint colors. As described in Section \ref{sec:directAssign}, merit for direct associations can be computed in multiple ways, but they simplify to the same outcome when there are two colors and two concepts \cite{schloss2018}. By limiting our simulations to the two endpoint colors and two endpoint concepts, we can think about merit for direct color-concept associations simply as association strength. This simplification assumes that colors between the endpoints vary monotonically in association strength with the domain concept (e.g., ocean water in Fig. \ref{fig:colormaps}).

\subsection{Relational associations as a source of merit}
\begin{figure}[tb]
 \centering
 \includegraphics[width=1.0 \linewidth]{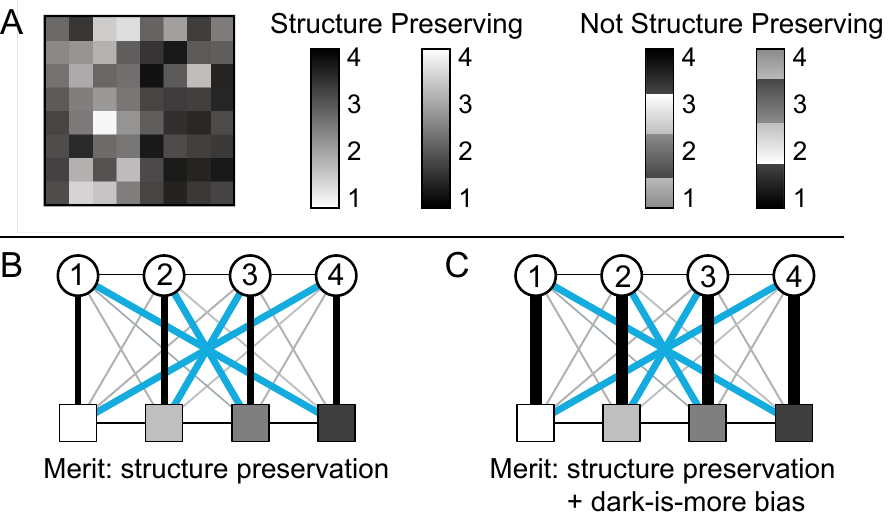}
 \caption{ Illustrations of structure preservation. (A) A colormap assigning lightness (light to dark) to quantity (1-4), with legends specifying structure-preserving assignments (natural progressions of lightness correspond to a natural ordering of quantities) vs. not structure-preserving assignments (assignment of lightness to quantity is scrambled). Bipartite graphs can code merit in terms of (B) structure preservation and (C) structure preservation plus the dark-is-more bias. 
 }
 \label{fig:structure_pres}
 \vspace{-6mm}
\end{figure}

To consider how relational associations can be represented as sources of merit in assignment inference for visualizations of continuous data, we first turn to Figs. \ref{fig:structure_pres}B-C. In these bipartite graphs, edges connect each possible color (shades from white to black) to each possible concept (numeric values from 1 to 4). As indicated in Fig. \ref{fig:structure_pres}B, only two possible sets of edges are structure-preserving with respect to the natural orderings of quantity and lightness: the set representing dark-more assignment (colored black) and the set representing light-more assignment (colored blue). Edges within each structure-preserving assignment receive more merit than edges that are not structure-preserving (colored gray), assuming that each set of structure-preserving edges is bound together (e.g., all blue or all black edges) and never a mix (e.g. some blue and some black edges). Based on structure preservation alone, dark-more and light-more assignments have equal merit, and thus should not be semantically discriminable.

Fig. \ref{fig:structure_pres}C shows merit from the combination of structure preservation and the dark-is-more bias. The dark-is-more bias places additional merit on structure-preserving edges representing dark-more assignment (thicker edges in Fig. \ref{fig:structure_pres}C). With greater merit on the dark-more assignment than light-more assignment, these two structure preserving assignments should be semantically discriminable. Given that treating the dark-is-more bias as a source of merit also implies structure preservation, here forward, we focus on the dark-is-more bias.

Fig. \ref{fig:colormaps} shows the dark-is-more bias represented as a source of merit for the example colormaps about ocean water. From the perspective of merit from the dark-is-more bias alone, assignment inference simulations for these colormaps predict that more ocean water should map to darker blues in the top row (consistent with direct associations) and darker browns in the bottom row (conflicting with direct associations). 
 
Like for direct associations, we reduced the problem to model only the endpoint colors and concepts. This simplification ensured that the edges from each potential structure-preserving set (dark-more and light-more) are not mis-matched during simulations of inferred mappings. Our approach assumes the colors in color scales used to construct colormaps vary monotonically in lightness, which was true in the present study (we return to this issue in the General Discussion).

\subsection{Combining direct and relational sources of merit}
We propose that assignment inference for visualizations of continuous data can be simulated using a weighted sum over multiple sources of merit. With knowledge on how much weight to put on merit from direct associations ($W_A$) and the dark-is-more bias ($W_D$), we can combine these sources of merit (combined merit bipartite graph in Fig. \ref{fig:colormaps}) and use established methods for simulating assignment inference \cite{schloss2018, schloss2021, mukherjee2022} to predict inferred mappings. In the top row, these sources of merit are consistent, and simulating assignment inference over combined merit predicts observers will infer that darker colors map to larger quantities. In the bottom row, these two sources of merit are conflicting. Depending on the relative weight given to each source, they might cancel each other out, or one might dominate over the other. The weights used in Fig. \ref{fig:colormaps} are based on the results of Exp. 3, with greater weight on direct associations than on the dark-is-more bias (see Exp. 3 for details).

In this study we asked whether direct and relational associations independently contribute to merit in assignment inference for colormap data visualizations, and if so, what is their relative contribution? Answering these questions enabled us to create a model that predicts people's inferred mappings, which can be used to help design colormaps that facilitate visual communication.

\section{Experiment 1}
Experiment 1 investigated whether both direct and relational color-concept associations contribute to inferred mappings for colormaps. We addressed this question using colormaps depicting fictitious data about two domain concepts, shade and sunshine. We chose these concepts because the dark-is-more bias and direct associations would be consistent for shade and conflicting for sunshine, allowing us to test for independent effects of each factor. 

\subsection{Methods}
We began by collecting direct color-concept association data for the domain concepts shade and sunshine. We then used these data to generate colormap stimuli to assess inferred mappings. Data, code, and color coordinates for all experiments in this paper can be found at https://github.com/SchlossVRL/assign-infer-colormaps.

\subsubsection{Measuring direct color-concept associations}\label{sec:Exp1assoc}
\begin{figure}[tb]
 \centering
 \includegraphics{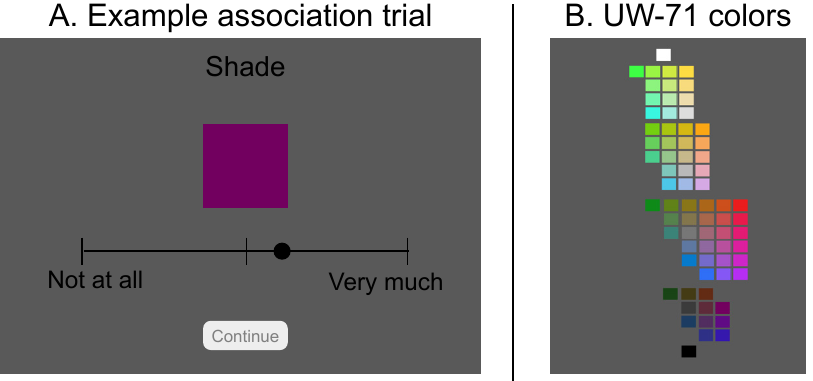}
 \caption{(A) Example association rating trial. The slider indicates a slight association for the given purple color with the concept shade. (B) The UW-71 colors as seen during the association task instructions.}
 \label{fig:assocTask}
 \vspace{-4mm}
\end{figure}

In the color-concept association task, participants were presented with a concept word at the top of the screen (sunshine or shade) and a colored square centered below (Fig. \ref{fig:assocTask}A). They rated how much they associated the given concept with the given color by moving a slider along a scale ranging from ``not at all'' ($-200$) to ``very much'' ($200$), and clicking ``continue'' to begin the next trial. Each concept was rated for each of the UW-71 colors \cite{mukherjee2022} shown in Fig. \ref{fig:assocTask}B (see Table \ref{table:UW_71_colors} in Supplementary Material for CIELAB coordinates). The UW-71 colors include 58 colors uniformly sampled from CIELAB space (UW-58 from \cite{rathore2020, schloss2021}), plus 13 additional colors sampled at a higher lightness plane to incorporate more saturated yellows and greens \cite{mukherjee2022}. 

Our target sample size was $n=30$ and we collected data from 35 Amazon mTurk workers given we expected several participants would be excluded for failing the attention check, described below (35 total, 3 excluded). The final sample was $n=32$ (mean age = 40 years old; 11 women, 21 men; gender assessed using free-response here and in all subsequent experiments). All participants indicated normal color vision when asked if they had difficulty distinguishing between colors relative to the average person and if they considered themselves colorblind. All participants of this and all subsequent experiments gave informed consent and the UW--Madison IRB approved the protocol.

Before beginning the task, participants were shown the domain concept words and all 71 colors (Fig. \ref{fig:assocTask}B). They were asked to identify which color they associated most and least with each concept to anchor the endpoints of the scale \cite{palmer2013}.The experiment was blocked by concept, with shade and sunshine presented in a random order within the first two blocks. The 71 colors appeared in a random order within each block. Given our plan to use association ratings from this task to generate stimuli for the colormaps task, we sought to include associations only from participants who made careful judgments. Thus, we included a third, attention check block for all participants and set an \textit{a priori} exclusion criterion (see Section \ref{sec:SuppAttnCheck} in the Supplementary Material). 

The displays of this and all subsequent experiments were created using jsPsych \cite{de2015jspsych}. All participants completed the experiments on their own devices so the color coordinates were calculated using standard assumptions about RGB displays. Thus, as is typical in color experiments in visualization, which aim to be robust to variations across displays \cite{stone2014, szafir2018, gramazio2017, mukherjee2022}, the precise colors each participant saw varied with the specifications of their monitors. This experiment took approximately 30 min.\ and participants were compensated \$3.63. The mean color-concept associations for sunshine and shade are shown in Fig. \ref{fig:SuppSunShadeHill} of the Supplementary Material.

\subsubsection{Generating colormaps}\label{sec:exp1map}
\begin{figure}[tb]
 \centering
 \includegraphics[width=1.0 \linewidth]{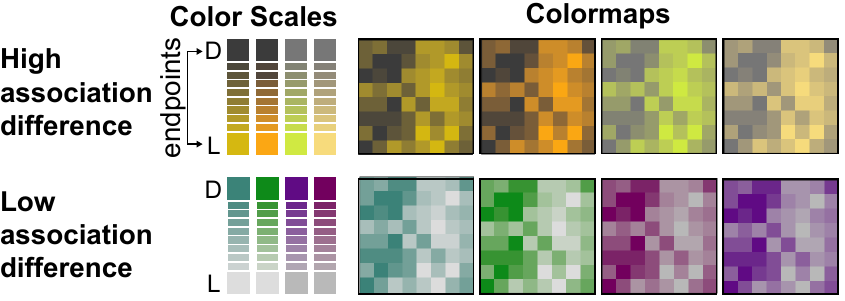}
 \caption{Color scales and corresponding colormaps in Exp. 1.}
 \label{fig:exp1methods}
 \vspace{-4mm}
\end{figure}

To generate the colormaps for this experiment, we (1) specified eight pairs of endpoint colors, (2) interpolated between the eight endpoint colors to create color scales, and then (3) applied the colorscales to 10 underlying datasets to create colormap data visualizations (Fig. \ref{fig:exp1methods}). 

(1) We selected endpoint colors such that one color was lighter (L) and the other was darker (D). For four endpoint pairs, association difference was high---sunshine was far more associated with the light than the dark endpoint, and shade was far more associated with the dark than the light endpoint. For four other pairs, association difference was low, which occurred when both colors were either weakly or moderately associated with the domain concept.\footnote{Overall, mean association ratings increased with lightness for sunshine (CIELAB L*) ($r(69) = .71, p < .001$) and decreased with L* for shade ($r(69) = -.79, p < .001$), but some light colors were moderately associated with sunshine, and some dark colors were moderately associated with shade. These properties enabled us to generate colormaps that varied in association difference.} Within each level of association difference, two endpoint pairs had a lightness difference of $L^*= 38$, and two had a difference of $L^* = 50$. We tested multiple color pairs for each condition to ensure our results were not specific to any one color pair. We checked if the colors interpolated between the endpoints varied (approximately) monotonically in direct association strength for both sunshine and shade (i.e., the domain concept was not more associated with the intermediate colors than with either endpoint)\footnote{Two color scales for shade did not meet our statistical criterion, due to a coding error treating hue angle as radians instead of degrees. However, our statistical criterion is a heuristic, and visual inspection suggested that the intermediate colors still varied monotonically between the endpoints (Supplementary Material Fig. \ref{fig:SuppMonoShadeSun}A), so we kept data for these color scales in the analysis.}. See Supplementary Material Section \ref{sec:SuppSelectColors} and Fig. \ref{fig:SuppMono} for details.

(2) Using these endpoint colors, we created eight color scales by linearly interpolating eight steps between the light and dark endpoints (interpolation computed in CIELAB space). The resulting color scales had 10 steps, as in the stimuli from \cite{schloss2019}.

(3) Finally, we applied each of the eight color scales to 10 underlying datasets, producing 80 colormap data visualizations. The underlying datasets produced colormaps appearing as an $8 \times 8$ grid, where one side was biased to be lighter and the other side was biased to be darker. Within the 10 underlying datasets, half produced colormaps in which the left side was darker than the right side (as in Fig. \ref{fig:exp1methods}), and the other half produced colormaps in which the right side was darker (as in Fig. \ref{fig:colormaps}). A full set of 10 colormaps from one color scale are shown in Fig. \ref{fig:SuppExp1colormaps} in the Supplementary Material. 

The underlying datasets we used were previously used to generate colormaps in \cite{schloss2019}. The data ranged from 0-1, with values sampled from eight discrete points along an arctangent curve with added noise. The eight points corresponded to the eight columns of the colormaps. The samples at each point were used to assign values to the rows within each column of the colormap (see Supplementary Material Section \ref{sec:SuppUnderData} for further details). One endpoint of the color scale was assigned a data value of 0 and the other endpoint a data value of 1, such that the color scales corresponded to the full range of the underlying data. Given that the data were evenly sampled along the arctangent curve, the data represented in the colormaps evenly span the full data range. This method of generating stimuli mitigates concerns about the dynamic range of data variability being hidden in the data visualization \cite{elmqvist2010, zeng2021}.

\subsubsection{Assessing inferred mappings for colormaps}\label{sec:exp1inf}
In the colormaps task, participants were presented with colormaps along with a domain concept (sunshine or shade). They were told that the colormaps represented amounts of sunshine (or shade) from various counties in a state. In some counties, there was more sunshine (shade) on the left side of the county; in other counties, there was more sunshine (shade) on the right side. Their task was to indicate whether there was more sunshine (shade) on the left/right of the map by pressing the left/right key on their keyboard.  

Domain concept and color scale varied between-subjects, and participants were randomly assigned to one of 16 groups (8 color scales $\times$ 2 domain concepts). Each participant judged all 10 colormaps for their assigned domain concept and color scale, one at a time in a random order.  Trials were separated by a 500-ms inter-trial interval. The colormaps (approx. 4cm $\times$ 4cm) appeared on a white square (approx. 9cm $\times$ 9cm) in the center of a medium gray screen (size estimates using a 15.6in, 1920 $\times$ 1080 pixel monitor). Below each half of the colormap was a horizontal line labeled ``Left''/``Right'' (Fig. \ref{fig:colormaps}). 

Our target sample size was $n=192$, $n=12$ per group (sample size based on a power analysis reported in Supplementary Material Section \ref{sec:SuppPower}). The final sample was 187 mTurk workers (mean age = 38.9 years old, 105 women, 82 men), after excluding $n=3$ for atypical color vision and $n=2$ for not completing the experiment. The groups ranged from $9-13$ participants due to how the experiment code automated assignments to conditions while managing exclusions. The experiment took approx. 5 min. and participants were compensated with \$0.60.

 \begin{figure}[htb]
 \centering
 \includegraphics{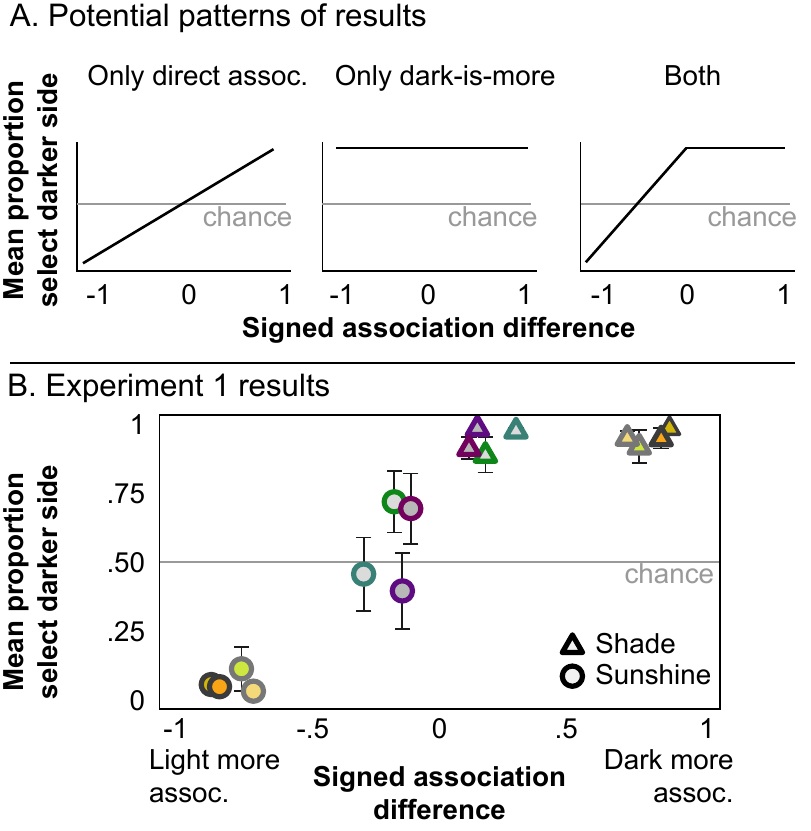}
 \caption{(A) Predicted patterns of results in Exp. 1 if inferred mappings are influenced by only the dark-is more bias (left), only direct associations (center), or both (right). (B) Results of Exp. 1, showing the mean proportion of times the darker side was selected for maps about sunshine (circles) and shade (triangles) as a function of signed association difference. Mark colors represent endpoint colors in the color scales and error bars represent standard errors of the means.}
 \label{fig:exp1results}
 \vspace{-4mm}
\end{figure}

\subsection{Results and Discussion}\label{sec:Exp1mapResults}
Fig. \ref{fig:exp1results}A shows potential patterns of inferred mappings if there was an effect only of direct associations, only of the dark-is-more bias, or both. The y-axis represents the proportion of times the darker side would be chosen over all trials, as a function of signed association difference. Positive/negative association differences indicate the darker/lighter color is more associated with the domain concept, respectively. Direct associations only predicts the probability of choosing the darker side would increase monotonically as the darker side becomes more associated with the domain concept. Dark-is-more only predicts participants would always choose the darker side, regardless of association difference. If both have an effect, then participants would choose the darker side when the two biases are consistent (positive association differences) but less likely to choose the darker side as the lighter side becomes more associated with the domain concept (negative association differences).

Fig. \ref{fig:exp1results}B shows the mean proportion of times the darker side was chosen, averaged over the 10 repetitions within each participant, and then averaged over participants. The pattern of responses resembles the predicted pattern if both the dark-is-more bias and direct associations influenced inferred mappings. Participants almost always chose the darker side for shade (association differences greater than zero), and their likelihood of choosing the darker side decreased as the lighter side became more associated with sunshine. 

To test for independent effects of each potential source of merit, we used a mixed-effect logistic regression model. Although we plot the data in terms of the proportion of times participants chose the darker side (Fig. \ref{fig:exp1results}B), this way of coding the data poses a problem for including the dark-is-more bias as a predictor in a regression model, given that there is no variability in the predictor (it predicts a response of `1' on every trial). Thus, we conducted a model to predict whether participants chose the left side on each trial (1 = left, 0 = right), from a predictor coding whether the left side was darker (1 = left darker, $-1$ = right darker), and a predictor coding which side was more associated, and by how much (scaled to range from $-1$ to 1; x-axis values in Fig. \ref{fig:exp1results}B). Conducting models with respect to the left side is a standard approach in psychophysics research, and is valid as long as the stimuli are left/right balanced, as in the present stimulus set (see Section \ref{sec:exp1map}).

Participants were more likely to select the left side if it was more strongly associated with the domain concept than the right side ($B = 4.51, SE = 0.21; z = 21.22, p < .001$) and if it was dark than light ($B = 1.33, SE = 0.09; z = 15.46, p < .001$) (dark-is-more bias). Thus, both direct and relational associations influenced inferred mappings. See Section \ref{sec:SuppAddAnalysis} in Supplementary Material for an additional analysis that includes concept as a factor in the model.

\textbf{Summary.} Exp. 1 showed that direct associations and the dark-is-more bias contribute independently to people's inferred mappings. When these two factors conflict (the domain concept is more associated with the light endpoint than the dark endpoint) and the direct association difference is large, direct associations override the dark-is-more bias.

\section{Experiment 2}\label{sec:Exp2assoc}
Given evidence that direct associations can override the dark-is-more bias when they conflict and direct associations are strong, we conducted Exp. 2 to test how much association difference was needed for direct associations to fully override the dark-is-more bias. The results led us to study effects of semantic distance for predicting inferred mappings.

 \subsection{Methods}\label{sec:Exp2mapMethods}
\begin{figure}[tb]
 \centering
 \includegraphics{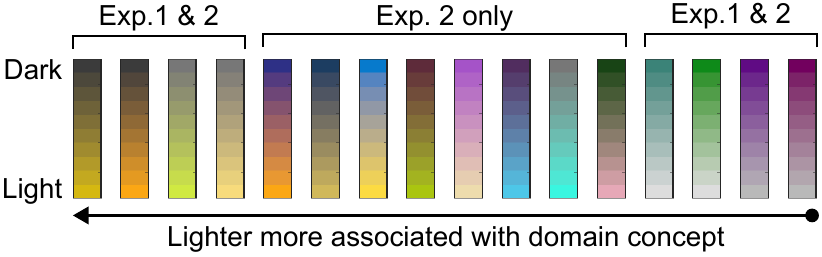}
 \caption{Color scales used in Exp. 2. From right to left, sunshine is increasingly more associated with the light endpoint of the color scale. The four leftmost and four rightmost color scales were used in Exp. 1.}
 \label{fig:exp2scales}
 \vspace{-1mm}
\end{figure}

The methods were the same as Exp. 1, except for two changes. First, we only tested sunshine as the domain concept in order to focus on cases where direct associations and the dark-is-more bias conflict. Second, we included eight new color scales of intermediate association difference, in addition to the original eight from Exp. 1 (16 color scales) (Fig. \ref{fig:exp2scales}).

For the new color scales, we selected endpoint color pairs using the association data from Exp. 1 with the same selection criteria (Supplementary Material Section \ref{sec:SuppSelectColors}). Two of the new color scales did not meet our statistical criterion for monotonicity due to a coding error treating hue angle as radians instead of degrees (Supplementary Material Fig. \ref{fig:SuppMonoShadeSun}B), and visual inspection showed that intermediate colors were more associated with sunshine than the endpoints. Thus, we excluded data from these two color scales from analysis. 

\looseness=-1

We focused on association difference and allowed association strength to vary (e.g., the same value of association difference could be achieved if sunshine was moderately associated with the light endpoint and weakly associated with the dark endpoint, or strongly associated with the light endpoint and moderately associated with the dark endpoint). As in Exp. 1, each participant judged 10 colormaps constructed from one of the 16 color scales (between-subjects). 

Our target sample size was $n=640$ ($n=40$ per condition) based on a power analysis (see Supplementary Material Section \ref{sec:SuppPower}). We collected data until each condition reached at least 40 participants after excluding those with atypical color vision (696 Amazon mTurk workers in total, 41 excluded). We assessed color vision using the two self-report questions in Exp. 1, plus responses to six digital Ishihara plates. Participants were excluded if they answered yes to either question and/or answered incorrectly for more than one of the six plates. The final sample included 655 participants (mean age = 38.6 years old; 294 women, 358 men, 2 non-binary, 1 no report). The groups ranged from $40-44$ participants due to how the experiment code automated assignments to conditions while managing exclusions. The experiment took approximately 5 min. and participants were compensated \$0.60.%

 \subsection{Results and Discussion}\label{sec:Exp2mapResults}
\begin{figure}[tb]
 \centering
 \includegraphics{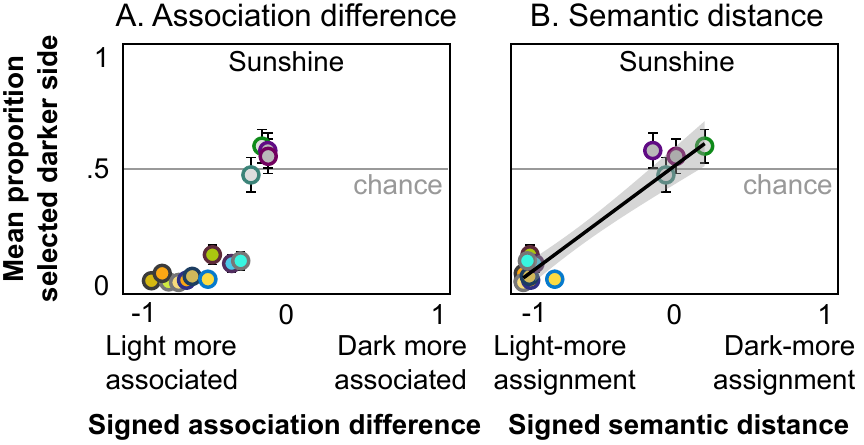}
  \vspace{-4mm}
 \caption{Mean proportion of times the darker side was selected for each color scale, plotted as a function of (A) signed association difference, and (B) signed semantic distance using direct associations as merit. Mark colors indicate endpoint colors. Error bars indicate standard error. }
 \label{fig:exp2results}
 \vspace{-5mm}
\end{figure}

Fig. \ref{fig:exp2results}A shows the mean proportion of times participants selected the darker side for colormaps generated from each color scale (averaged over the 10 colormaps judged by each participant, and then averaged over participants).  As in Exp. 1, direct associations were more likely to override the dark-is-more bias as association difference increased ($r(12) = .85, p < .001$). But, once association difference reached about $-.55$, participants almost always inferred that the lighter side of the colormaps mapped to more sunshine. Direct associations fully overrode the dark-is-more bias, so further increasing association difference could not further influence inferred mappings (floor effect). This observation led us to ask, why would inferred mappings level off at around $-.55$?

One possibility is that participants approached this task using assignment inference, comparing each possible assignment (dark-more or light-more), and inferring the assignment with greater merit. Once the merit of one assignment is sufficiently greater than the alternative, the colors reach maximal semantic discriminability. Then, further increasing association difference has no further effect on assignment inference. This limit may have been reached at an association difference of around $-.55$. If so, then the plateauing function in Fig. \ref{fig:exp2results}A may become linear when we replace the x-axis (signed association difference) with simulations of assignment inference (Section \ref{sec:directAssign}).

To simulate assignment inference for each color scale, we first calculated semantic distance for each pair of endpoint colors (using equation \ref{eq:sd} defined in \cite{schloss2021} and reproduced in Supplementary Material Section \ref{sec:SuppSemDist} of this paper). We then determined the optimal assignment (i.e., which assignment had greater merit), and coded the outcome as dark-more = $+1$ and light-more = $-1$. Last, we multiplied this coding by semantic distance to compute \textit{\textbf{signed semantic distance}}, which gave positive values to the probability of inferring dark-more assignments and negative values to the probability of inferring light-more assignments.

Computing signed semantic distance required specifying merit based on direct associations between each endpoint color and each endpoint concept (``no sunshine'' vs. ``a lot of sunshine''), as in Fig. \ref{fig:colormaps}. From Exp. 1, we had association data for domain concept ``sunshine,'' but not the endpoint concepts. Thus, we collected data from additional participants ($n = 31$), who rated the association strength between each endpoint color and endpoint concepts ``no sunshine'' and ``a lot of sunshine'' (see  Supplementary Material Section \ref{sec:SuppAddData}). We used the mean ratings as merit to compute signed semantic distance. 

As shown in Fig. \ref{fig:exp2results}B, inferred mappings were predicted by simulations of assignment inference:  
signed semantic distance was strongly correlated with the proportion of times participants chose the darker side ($r(12) = .97, p < .001$). This correlation was stronger than the correlation for signed association difference reported above ($z = 1.96, p = .05$). The trail of points that plateaued in Fig. \ref{fig:exp2results}A now compress onto signed semantic distance values near $-1$ in Fig. \ref{fig:exp2results}B.

From this strong linear relation, one may suppose that only merit from direct associations is needed to simulate inferred mappings for colormap visualizations. But, Exp. 2 only included conditions in which direct associations and the dark-is-more bias \textit{conflicted}, and Exp. 1 suggested that when they were \textit{consistent}, the dark-is-more bias dominated regardless of association strength difference. To understand the relative contribution of these two potential sources of merit in assignment inference, it is necessary to model data sampled from multiple points along the full range of signed semantic distance (see Exp. 3).

\textbf{Summary.} As in Exp. 1, Exp. 2 showed that direct associations override the dark-is-more bias when the association difference between the light and dark colors was sufficiently large. The pattern of inferred mappings was strongly predicted by simulations of assignment inference using merit from direct associations (signed semantic distance).

\section{Experiment 3}\label{sec:Exp3}
In Exp. 3, we developed and tested a new method to combine multiple (sometimes conflicting) sources of merit to simulate assignment inference (Fig. \ref{fig:colormaps}). The experimental task was the  same as Exps. 1-2, but we tested three new domain concepts (ocean water, wild fire, and glacial ice), and sampled 21 points along the full range of direct association-based signed semantic distance for each concept. Our goals were: (1) determine the optimal weighting on direct associations ($W_A$) relative to the dark-is-more bias ($W_D$) when computing combined merit, and (2) test whether simulations of assignment inference using the optimal combined merit predicted people's inferred mappings better than simulations using each source of merit alone. 

\subsection{Methods}
We first collected direct association ratings for the domain concepts and then used the mean ratings to generate colormaps to assess inferred mappings. We also collected additional data to quantify merit for direct associations and the dark-is-more bias. 

\subsubsection{Measuring direct color-concept associations}

We collected direct association ratings for five domain concepts relevant to environmental data: wild fire, ocean water, glacial ice, ground soil, and tree foliage (inspired by \cite{samsel2017}), using the same methods as in Exp. 1. The data are shown in Supplementary Material Fig. \ref{fig:SuppEnvironhill}.

The target sample was $n=35$ to match Exp. 1, and we collected data in batches until reaching this target after excluding those with atypical color vision ($n=15$) and who failed the attention check ($n=17$); 70 Amazon mTurk workers total. We shortened the attention check block but used the same \textit{a priori} exclusion criterion (see Supplementary Material Section \ref{sec:SuppAttnCheck}). Our final sample was $n=38$ (mean age = 42.8 years old, 18 women, 20 men). The experiment took approximately 60 min. and participants were compensated \$7.25.

\subsubsection{Generating colormaps and computing merit} \label{sec:exp3_maps_merit}
\begin{figure}[tb]
\centering
   \includegraphics{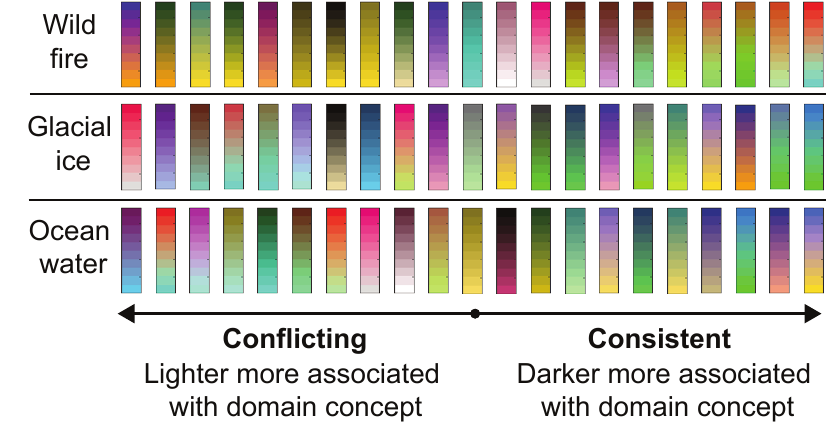}
\caption{Color scales used to create colormaps in Exp. 3. Rightward of center, the domain concept increases in direct association strength with the \textit{darker} endpoint (consistent direct and relational associations). Leftward of center, the domain concept becomes increasingly more associated with \textit{lighter} (conflicting direct and relational associations).}
 \label{fig:exp3colorscales}
 \vspace{-5mm}
\end{figure}

Based on the direct association data, we created colormaps for three concepts: wild fire, ocean water, and glacial ice. These domain concepts enabled spanning the full range of signed semantic distances within each concept. We generated colormaps using the methods in Exp. 1 (Fig. \ref{fig:exp1methods}) and Supplementary Material Section \ref{sec:SuppSelectColors}. For each domain concept, we chose 21 pairs of endpoint colors that spanned the full range of association differences from strongly negative (light color was more associated with the domain concept) to strongly positive (dark color was more associated with the domain concept). Fig. \ref{fig:exp3colorscales} shows the resulting 21 color scales for each domain concept. All color scales satisfied the criterion for monotonicity. As in Exps. 1 and 2, each color scale was applied to 10 underlying datasets to produce 10 colormaps per color scale, with darker side left/right balanced (Fig. \ref{fig:SuppExp1colormaps}).

After selecting the color pairs, we collected additional data to estimate merit for each endpoint color paired with each endpoint concept, with respect to direct associations and the dark-is-more bias (Fig. \ref{fig:colormaps}).

\textit{\textbf{Merit for direct associations.}} A new set of 30 participants rated the association strength between each endpoint of each domain concept (e.g., ``a lot of ocean water,'' ``no ocean water'') and each corresponding endpoint color (details in Supplementary Material Section \ref{sec:SuppAddData} ). As in Exp. 2, we used these associations to estimate merit derived from direct associations for each color-concept endpoint pairing  (Fig. \ref{fig:colormaps}).

\textit{\textbf{Merit for the dark-is-more bias.}} So far, we have discussed the dark-is-more bias as binary---dark-more assignments have greater merit than light-more assignments. However, the dark-is-more bias can also be considered continuous---the degree to which dark-more assignments have greater merit depends on the degree to which one endpoint appears clearly darker than the other endpoint. One might consider quantifying merit of the dark-is-more bias using lightness (L*) difference between the two endpoint colors of the color scale that varied monotonically in lightness. However, we reasoned that the dark-is-more bias would be activated if one side appeared clearly darker than the other, and adding additional lightness difference may not increase activation of the bias. 

Thus, we used an empirical approach to quantify merit for the dark-is-more bias. For each endpoint color pair, volunteers with expertise in color perception ($n=4$) rated the degree to which one color was clearly darker than the other color (referred to as \textit{darkness difference ratings}). They judged each pair twice (left/right balanced), and made their ratings on continuous slider scale from ``left color is clearly darker'' to ``right color is clearly darker.'' The middle was  labeled ``equal darkness'' (see Supplementary Material Section \ref{sec:SuppAddData} for details). For each color scale, we coded dark-more edges to have merit = 1 and light-more edges to have merit = 0, and then multiplied these values by the darkness difference ratings. As a result, differences in total merit of dark-more vs. light-more assignments scaled with the degree to which it was obvious that the dark endpoint appeared darker than the light endpoint.

\subsubsection{Colormap interpretation task }\label{sec:Exp3map}
This task was the same as in Exps. 1 and 2, except the domain concepts were wild fire, ocean water, and glacial ice and there were 21 color scales per domain concept (3 domain concepts $\times$ 21 color scales = 63 groups of participants). Participant judged 10 colormaps for their assigned domain concept and color scale. They were told that the colormaps represented data about [domain concept] in different counties. Their task was to indicate whether there was more [domain concept] on the left/right side of the county (Fig. \ref{fig:colormaps}).

The target sample size was $n=1260$ ($n=20$ per condition) based on a power analysis (see Supplementary Material Section \ref{sec:SuppPower}). We collected data in batches until each condition had at least $n=20$ after excluding those with atypical color vision as assessed in Exp. 2 (1391 Amazon mTurk workers total, 107 excluded). The final sample was $n=1284$ (mean age = 40.3 years old, 1 no reported age; 672 women, 598 men, 9 non-binary, 5 no reported gender). The groups ranged from $20-22$ participants due to how the experiment code automated assignments to conditions while managing exclusions. The experiment took approximately 5 min. and participants were compensated \$0.60.

\subsection{Results and Discussion}\label{sec:Exp3Results}
In the following analyses, we determined the optimal relative weighting on direct and relational associations, and then assessed whether assignment inference simulations using the optimal weighting predicted inferred mappings better than simulations using each source of merit alone. We split participants into a training set to determine the optimal weighting, and a test set to compare the optimal weighting with each source of merit alone. Each set had $10-12$ participants per color scale for each domain concept. We simulated assignment inference with varying relative weight on each source of merit as follows: 

\textbf{(1) Computing combined merit.} First, we specified merit of each color-concept pairing within each source of merit (Section \ref{sec:exp3_maps_merit}). Then, we calculated combined merit by computing the weighted sum over bipartite graphs for each source of merit (Fig. \ref{fig:colormaps}). We used each combination of weights on direct associations ($W_A$) and the dark-is-more bias ($W_D$) in increments of .05, such that their sum was 1. Each weight was a multiplicative factor on each edge of the respective bipartite graphs. For instance, a weight pairing of (1,0) placed all the weight on direct associations, (0,1) placed all the weight on the dark-is-more bias, and (.5, .5) placed equal weight on both sources of merit.

\textbf{(2) Computing signed semantic distance.} We computed signed semantic distance over combined merit for each weight pairing, following the procedure in Exp. 2. First, we computed semantic distance between the endpoint colors for each domain concept. Next, we determined the optimal assignment (which assignment had greater overall merit), coded as $+1$ for dark-more and $-1$ for light-more. Last, we multiplied this coding by semantic distance to obtain signed semantic distance.   

\begin{figure}[h!!]
\centering
   \includegraphics[width=1.0\linewidth]{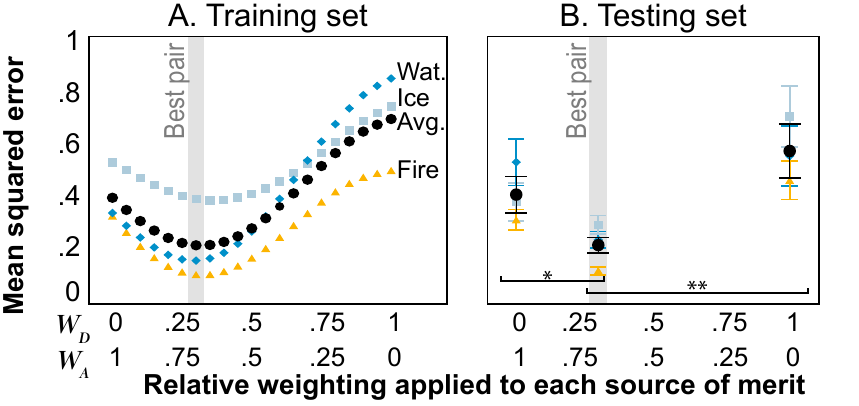}
\caption{Mean squared error (MSE) predicting inferred mappings from assignment simulations with varying weights on dark-is-more bias ($W_D$) and direct associations ($W_A$) for the (A) training set and (B) testing set. MSEs are shown separately for colormaps representing ocean water (blue diamonds), glacial ice (gray squares), and wildfire (yellow triangles), plus the average of all three concepts (black circles). Error bars represent standard error of the means. The gray bar indicates the best pair, determined from the training set.}
\vspace{-5mm}
 \label{fig:exp3mseresults}
\end{figure}

To determine the optimal weighting, we used mean squared error (MSE) to compare assignment inference simulations with human judgments. For each of the 21 color scales for each of the three domain concepts, we computed MSE between signed semantic distance and the mean probability that participants in the training set chose the darker side of the colormaps. When computing MSE, we scaled the proportion chosen data to range from -1 to 1, corresponding with the scale of signed semantic distance. Fig. \ref{fig:exp3mseresults}A shows MSEs averaged over the 21 color scales for each domain concept, plotted as a function of weight pairs, along with the average over domain concepts. On average, the best performing weight pair yielding the lowest MSE had a weight of $W_A = .7$ on direct associations and $W_D = .3$ on the dark-is-more bias.

Using data from the held out testing set, we evaluated whether this optimal weight pair was better for predicting assignment inference than each source of merit alone. For each color scale for each domain concept, we computed MSE between mean responses (scaled to range from $-1$ to $1$) and signed semantic distance with the optimal weighting identified from the training set (.7, .3), with all weight on direct associations (1,0), and with all weight on the dark-is-more bias (0,1) (Fig. \ref{fig:exp3mseresults}B). To test effects of relative weighting, we used a linear mixed effects model predicting MSE for each color scale, with fixed effects for relative weighting, domain concept, and their interaction (using Helmert contrasts). The model also included a by-color scale random intercept and random slope for relative weighting. Here we focus on the main effect of relative weighting ($F(2,53.45) = 9.18, p <.001$), and we report on further details of this model in Supplementary Material Section \ref{sec:SuppAddAnalysis}. Planned independent samples t-tests indicated that the optimal weight pair fit inferred mappings better than direct associations alone ($t(124) = -2.55, p = .01$) and dark-is-more alone ($t(124) = -3.53, p = .001$).
 
\begin{figure}[h!]
\centering
   \includegraphics[width=1.0\linewidth]{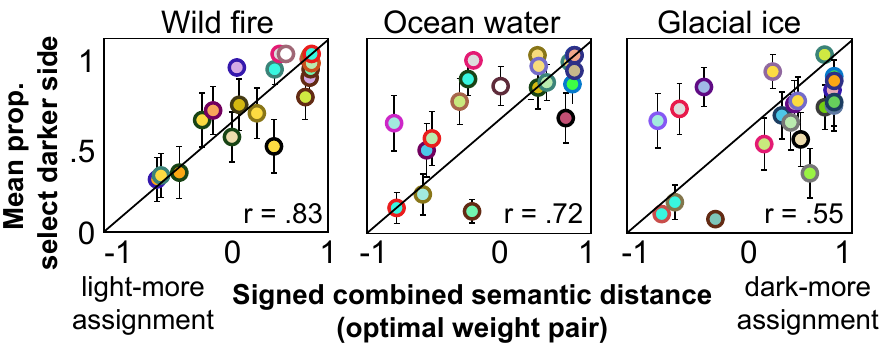}
   \vspace{-5mm}
\caption{Mean proportion of times the darker side was selected for the held-out testing set for each color scale as a function signed semantic distance using the optimal weight pair. Mark colors indicate color scale endpoint colors, and error bars indicate standard error.}
 \label{fig:exp3bestWeights}
\end{figure}

 Fig. \ref{fig:exp3bestWeights} shows the relation between participant responses and simulations of assignment inference using the optimal weight pairings for each color scale. Points would fall along the diagonal line if the simulations perfectly predicted inferred mappings. Signed semantic distance was significantly correlated with inferred mappings for all three domain concepts,  but to varying degrees: strong correlation for wild fire ($r = .83, p < .001$), moderately strong for ocean water ($r = .72, p < .001$), and moderate for glacial ice ($r = .55, p = .01$). Preliminary exploration suggests this weaker relation for glacial ice might be due to some colormaps appearing to vary in opacity, activating the opaque-is-more bias. The opaque-is-more bias aligns with the dark-is-more bias on light backgrounds (as used here), and the two relational associations may have combined to jointly override effects of direct associations. Our study was not designed to test the opaque-is-more bias, so future work is needed to study these effects more directly.\footnote{This weaker correlation for glacial ice is not likely due to concept \textit{glacial ice}, but rather the colormaps appearing to vary in opacity happened to be in the glacial ice condition. The applicability of the opaque-is-more bias depends on the combination of background and colors of the color scale \cite{schloss2019}, and would be applicable for these colormaps if they represented any other domain concept.}

Supplementary Material Section \ref{sec:SuppAddAnalysis} includes an additional analysis showing independent effects of direct and relational associations on inferred mappings (as in Exp. 1). It also includes plots of inferred mappings as a function of association difference and semantic distance from direct associations (Fig. \ref{fig:SuppExp3assocDist}), analogous to Fig. \ref{fig:exp2results} in Exp. 2. The strong plateau for sunshine colormaps in Exp. 2 was less apparent in Exp. 3, and we consider possible explanations in Section \ref{sec:Exp3SupResults}.

\textbf{Summary.} Exp. 3 showed that inferred mappings for colormaps were well-predicted using a simulation of assignment inference with combined merit. The optimal combined weighting resulted in predictions with less error than predictions simulated with weight on direct associations or the dark-is-more bias alone.

\section{General discussion}\label{sec:genDiscuss}
A central problem in visual communication is understanding how people infer meaning from visual features. By anticipating people's expectations about how visual features should map onto concepts, designers can create visualizations that align with those expectations, thereby facilitating communication \cite{norman2013, tversky2002, tversky2011, hegarty2011, lin2013, schloss2018, schloss2019, schloss2021, sibrel2020, mukherjee2022}. 

We approached this problem by bridging work on inferred mappings for visualizations of categorical information \cite{lin2013, setlur2016, schloss2018, schloss2021, mukherjee2022} and visualizations of continuous data \cite{cuff1973, mcgranaghan1989, schloss2019, sibrel2020} to understand both within a single framework of assignment inference. Doing so required broadening the notion of merit in assignment inference to include not only direct associations as in \cite{schloss2018, schloss2021, mukherjee2022}, but also relational associations (e.g., dark-is-more bias). Exp. 1 showed that direct and relational associations contribute independently to inferred mappings for colormaps. Exp. 2 showed that inferred mappings for colormaps were predicted by simulations of assignment inference (signed semantic distance) using merit from direct associations. Exp. 3 showed that simulating assignment inference using a weighted sum over merit from direct and relational associations was better at predicting inferred mappings than simulations using each source of merit alone.   

This study is an initial step towards comprehensively modeling the effects of multiple sources of merit in assignment inference. Here, we began with direct associations and one type of relational association, the dark-is-more bias. In future work, we will extend our approach to include additional sources of merit, including the opaque-is-more bias and hotspot-is-more bias. To quantify merit for the opaque-is-more bias, it will be necessary to estimate the degree to which colors in the colormap appear to vary in opacity depending on the background color (see \cite{schloss2019}), and ensure that this source of merit falls out of the equation when colormaps do not appear to vary in opacity. To quantify merit for the hotspot-is-more bias, it will be necessary to quantify the degree to which hotspots are salient in the colormap, and again ensure that this source of merit falls out of the equation when colormaps do not appear to have hotspots. Our approach for estimating inferred mappings not only has potential to accommodate known sources of merit, but can also scale as additional sources of merit are discovered. 

We also expect our approach to extend to abstract concepts. Evidence suggests that sets of abstract concepts previously considered ``non-colorable'' (e.g., sleeping, driving, safety, comfort) can be meaningfully encoded using color as long as their association distributions are sufficiently different from one another (semantic discriminability theory \cite{mukherjee2022}). In the present framework, as long as the colors in the color scale vary in association strength with the domain concept, then merit from direct associations will influence combined merit with the dark-is-more bias. If the associations do not vary in association strength (low semantic distance), then merit from direct associations will have little effect on combined merit, and the dark-is-more bias should dominate inferred mappings. These patterns should hold regardless of whether the concepts are abstract/concrete. If a concept has no systematic color-concept associations, regardless of whether it is abstract/concrete, then it will not be possible to create a color scale with large direct association difference, so the dark-is-more bias (and any other sources of merit) would dominate inferred mappings. 

Overall, our findings can be translated to incorporate color semantics into tools that generate colors for information visualizations (e.g., Colorgorical \cite{gramazio2017}, Color Crafter \cite{smart2019}, and CCC-Tool \cite{nardini2021}). These tools already allow designers to balance different factors, such as perceptual discriminability and aesthetics. With a comprehensive model of assignment inference combining multiple sources of merit, it will be possible to incorporate semantic discriminability into algorithms that optimize color selection for visualization design.  

\textbf{Limitations.} This study has limitations for future work to address.

\textit{\textbf{Linearly interpolated color scales.}} We used color scales that were linearly interpolated between two endpoints in CIELAB space, which supported the goals of this study. Interpolated color scales allowed us to compare merit of dark-more vs. light-more assignments using direct color-concept associations from only the endpoint colors. Using only the endpoints was possible because the intermediate colors varied approximately monotonically in association strength between the endpoint colors (see Supplementary Material Section \ref{sec:SuppSelectColors}). Monotonicity would be violated if the domain concept was more/less associated with intermediate colors of a color scale than the endpoints (e.g., using a color scale for sunshine that interpolated between a red and yellowish-green, resulting in more strongly associated yellows in the middle). 

Monotonicity would also likely be violated in industry standard color scales that spiral through color space \cite{smart2019, brewer1997, kindlmann2002}. Yet, Smart et al. \cite{smart2019} showed that such color scales that spiral produce colormaps that are more interpretable and aesthetically preferable than linear colormaps like the ones in the present study. Indeed, many criteria determine whether color scales (also referred to as ramps) are effective for visualizing continuous data \cite{samsel2017, bujack2018, zhou2016, silva2011, smart2019, rheingans2000, nardini2021} and our color scales were not designed to meet those criteria. Thus, the color scales in the present study are not meant to be used for visualizations of real data. To apply our modeling approach to more complex color scales, it will be necessary to quantify merit for color-concept pairings sampled in multiple steps between the two endpoints, and use a method for computing the optimal assignment that accounts for many colors and many concepts.

\textbf{\textit{Sequential color scales.}} The present work, and most previous work on inferred mapping for colormaps \cite{schloss2019,sibrel2020, mcgranaghan1989}, has focused on sequential color scales, where encoded data ranged from small to large. Questions remain concerning how this work extends to diverging scales, where encoded data has a neutral point. The dark-is-more and opaque-is-more biases imply that more extreme data (furthest from the neutral point) should map to darker, more opaque regions, respectively. Future work is needed to test these hypotheses, and to investigate people's expectations about which colors represent data values above/below the neutral point. Future work is also needed to determine whether the relative weightings on sources of merit established in Exp. 3 for sequential color scales generalize to diverging color scales. 

\textbf{\textit{Task type.}} Our study and much of the previous studies on inferred mappings for colormaps \cite{schloss2019,sibrel2020, mcgranaghan1989} used tasks that asked participants to interpret where ``more'' was represented in a colormap. However, people use colormap data visualizations for a wide variety of other tasks, such as those studied by Padilla et al. \cite{padilla2016}: finding a specific value of a concept, comparing values across regions, and averaging values across regions. Future research is needed to test whether the present framework modeling combined merit to estimate inferred mappings predicts performance in these other kinds of tasks. 

\textbf{Conclusion.} This work builds a new bridge for understanding how direct and relational associations combine to influence inferences about the meanings of colors in visualizations. We have laid the groundwork to develop a more comprehensive model of assignment inference that accounts for additional sources of merit that we know of, and can scale to accommodate new sources of merit as they are discovered. Our findings can be translated directly to design visualizations that align with people's expectations about the meanings of colors, thereby making visualizations that are easy to interpret.

\acknowledgments{
We thank Kushin Mukherjee, Chris Thorstenson, Anna Bartel, Clementine Zimnicki, Madeline Parker, and Audrey Wang for helpful discussions on this work. This project was supported by the UW--Madison Office of the Vice Chancellor for Research and Graduate Education, Wisconsin Alumni Research Foundation, NSF award BCS-1945303 to KBS, and NSF award $\#$2050782.}


\clearpage

\bibliographystyle{abbrv}

\bibliography{Sunshine-Full-vis.bib}


\clearpage
\appendix
\renewcommand*{\thesection}{S}
\counterwithin{figure}{section}
\counterwithin{table}{section}

\section{Supplementary Material}\label{sec:supplementary}

 \subsection{Underlying data distribution for generating colormaps}\label{sec:SuppUnderData}
 The following description of the process for generating the colormaps, including Fig.  \ref{fig:SuppUnderData}, is reproduced directly from pp. 813-814 in \cite{schloss2019}: 

\begin{adjustwidth}{1em}{2.5em}
The data used to generate the colormaps were sampled from an arctangent curve with added normally-sampled noise (Fig. \ref{fig:SuppUnderData}). To generate the data for each row of the colormap, we discretized the arc tangent curve into eight bins,corresponding to the eight columns in the colormap display. We centered the arctangent curve between the fourth and fifth bins, such that half of the display was biased to have larger values than the other half. We then perturbed each arctangent value by sampling from a normal distribution with the mean equal to the arctangent value and the standard deviation equal to 0.25. When the values fell outside the [0,1] interval, we re-sampled until they were all within the correct range. For half of the datasets, the arctangent curve was oriented as shown in Fig. \ref{fig:SuppUnderData}, and for the other half, it was left/right reversed. This enabled a left/right balance of the darker region (i.e., half of the colormaps contained the darker region on the left and the other half contained the darker region on the right).

 \begin{figure}[h]
 \centering
 \includegraphics{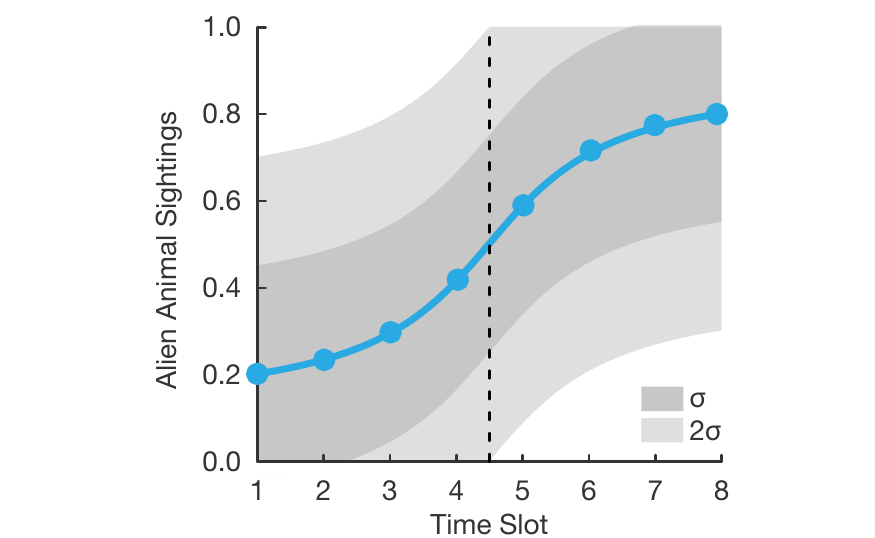}
 \caption{Distribution used to sample values at each time point to generate the data used to construct the colormap images. Figure reproduced from Figure 4 in Schloss et al. \cite{schloss2019}.}
 \label{fig:SuppUnderData}
 \vspace{-4mm}
\end{figure}

\end{adjustwidth}

 \subsection{Calculating semantic distance}\label{sec:SuppSemDist}
 We calculated semantic distance as specified in \cite{schloss2021, mukherjee2022}. The following definition is reproduced directly from p. 700 in \cite{mukherjee2022}:

\begin{adjustwidth}{1em}{2.5em}

  \textbf{Semantic distance} is a way to operationalize semantic discriminability in the case where there are $n=2$ features and concepts~\cite{schloss2021}. Fig.~\ref{fig:semdisteq} illustrates an example in which we have concepts \{M,W\} and colors \{1,2\}. The color-concept associations between all possible pairs are $x_1,\dots, x_4$, as shown in Fig.~\ref{fig:semdisteq}. We assume each $x_k$ is normally distributed with mean $\bar x_k$ equal to the corresponding $a_{ij}$ and standard deviation $\sigma_k = 1.4\cdot\bar x_k(1-\bar x_k)$, which was found to be a good fit to experimental data~\cite{schloss2021}. The outcome of the assignment problem is determined by the quantity $\Delta x := x_1-x_2+x_3-x_4$. The optimal assignment is:
    (M-1 and W-2 if $\Delta x > 0$) and (M-2 and W-1 if $\Delta x < 0$).

    Semantic distance is defined by the equation
    \begin{equation}\label{eq:sd}
        \Delta S = \bigl|\text{Prob}(\Delta x > 0)-\text{Prob}(\Delta x < 0)\bigr|.
    \end{equation}
    Since the $x_k$ are assumed to be normally distributed, so is $\Delta x$, and the probabilities in \eqref{eq:sd} can be computed analytically:
    \begin{equation}\label{eq:probpositive}
        \text{Prob}(\Delta x >0) = \Phi\Biggl(\frac{(\overline{x}_1+\overline{x}_4)-(\overline{x}_2+\overline{x}_3)}{\sqrt{\sigma_1^2+\sigma_2^2+\sigma_3^2+\sigma_4^2}} \Biggr),
    \end{equation}
    and $\text{Prob}(\Delta x < 0)=1-\text{Prob}(\Delta x > 0)$, where $\Phi(\cdot)$ is the cumulative distribution function (cdf) of the standard normal distribution.
    When $\Delta S$ is close to $0$, $\Delta x$ has a similar probability of being positive or negative, so the assignment is fragile. When $\Delta S$ is close to $1$, $\Delta x$ is almost always positive or almost always negative, so the assignment is robust. This notion of semantic distance can be used even when the features are not colors, by replacing the color-concept associations with feature-concept associations, and adjusting the formula for $\sigma_k$ as appropriate. 
    
    \begin{figure}[ht]
     \centering
     \includegraphics[width=1.0\columnwidth]{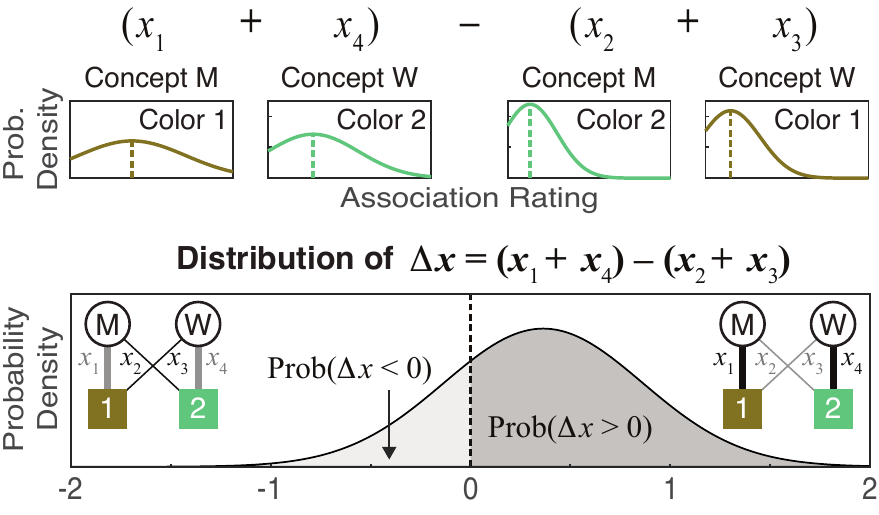}
     \caption{Diagram showing how association ratings between concepts \{M,W\} and colors \{1,2\} produce a distribution for $\Delta x$. Semantic distance is the absolute difference of the area under the curve to the left and right of zero. Figure reproduced from Figure 3 in Mukherjee et al.~\cite{mukherjee2022}, which was adapted from Figure 4 in Schloss et al.~\cite{schloss2021}.}
     \label{fig:semdisteq}
     \vspace{-2mm}
    \end{figure}
 \end{adjustwidth}
 
 \subsection{Attention check in color-concept association task}\label{sec:SuppAttnCheck}
\begin{figure}[ht]
     \centering
     \includegraphics{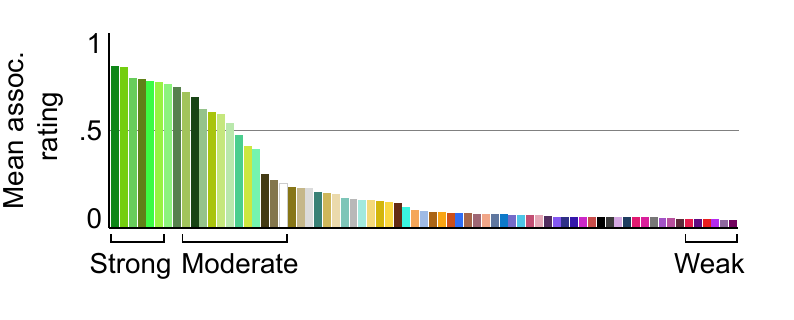}
     \caption{Mean association ratings for celery from Mukherjee et al. \cite{mukherjee2022}. The six strongest and six weakest colors were used as the basis for attention check exclusion criteria in Experiments 1 and 3 of the present work. In Exp. 3, only these twelve colors and the twelve moderate colors were shown during the association ratings task attention check.}
     \label{fig:celery}
    \end{figure}

 At the end of color-concept association tasks in Exps. 1 and 3, we included an additional block of trials to serve as an attention check. The goal was to collect color-concept association data for a concept for which participants should have strong, specific associations if they were making careful ratings during this task. We chose the concept celery, because we knew from previous data \cite{mukherjee2022} that participants (on average) had strong specific color-concept associations for celary (Fig. \ref{fig:celery}). That is, shades of greens were strongly associated with celery, and shades of reds and purples were weakly associated with celery. Using data from \cite{mukherjee2022}, we identified the six strongest and six weakest associated colors with celery (See Fig. \ref{fig:celery}). We then defined an \textit{a priori} exclusion criterion that the participants of the present study must have an association rating greater than .5 (on a scale from 0 to 1) for at least 5 out of 6 strongest celery colors, and an association rating less than .5 for at least 5 out of 6 weakest celery colors. We chose to put the attention check as the last block to evaluate performance at the end of the task when participants were most fatigued.

 In Exp. 1, participants rated the association between each of the UW-71 colors and the concept celery. In Exp. 3 we reduced the length of the attention check by having participants rate a subset of colors, rather than all 71 colors. We included the six colors most-associated celery colors, the six least-associated colors, and 12 moderately associated colors that straddled the middle of the ratings scale (around .5), determined from data in \cite{mukherjee2022} shown in Fig. \ref{fig:celery}.

 \subsection{Selecting endpoint colors to create color scales}\label{sec:SuppSelectColors}
\begin{figure}[h]
 \centering
 \includegraphics{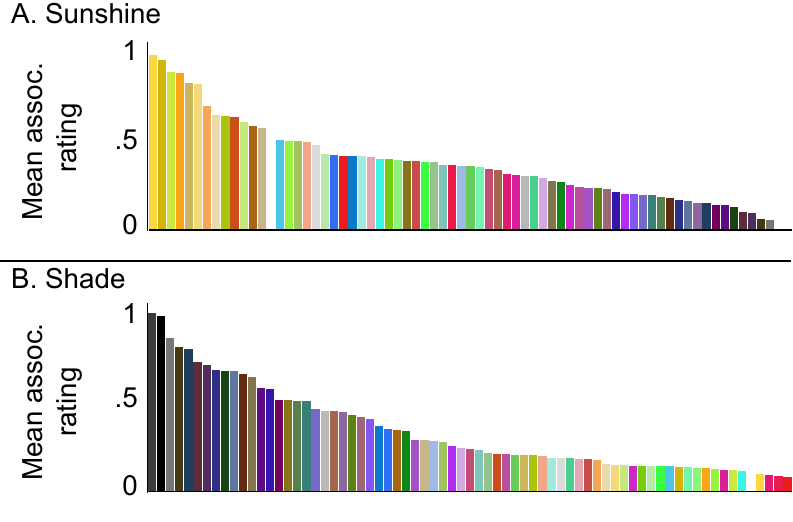}
 \caption{Mean direct association ratings for (A) sunshine and (B) shade with each of the UW-71 colors (fill of bar). Colors are sorted from high to low association strength.}
 \label{fig:SuppSunShadeHill}
\end{figure}

\begin{figure}[h!]
 \centering
 \includegraphics{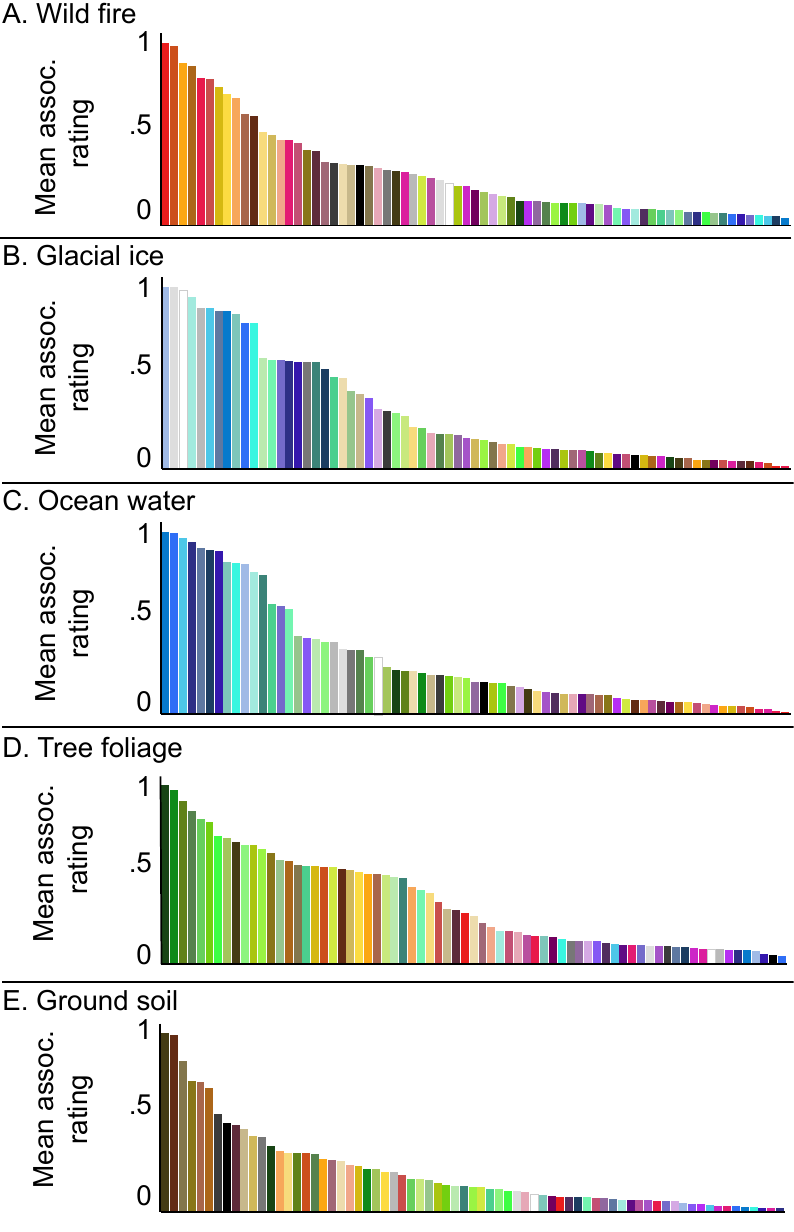}
 \caption{Mean association ratings for (A) wild fire, (B) glacial ice, (C) ocean water, (D) tree foliage, and (E) ground soil with each of the UW-71 colors (fill of bar). Colors are sorted from high to low association strength.}
 \label{fig:SuppEnvironhill}
 \vspace{-4mm}
\end{figure}

 In all three experiments, we used the same general process for selecting pairs of colors to serve as endpoints for creating color scales (Figs. \ref{fig:SuppSunShadeHill} and \ref{fig:SuppEnvironhill}). Below, we explain this approach in detail for Exp. 1, and explain any changes to this approach for Exps. 2 and 3. 
 
\textbf{Experiment 1.} To generate the color scales, we needed pairs of endpoint colors that varied in lightness and varied in their direct associations with shade and sunshine. We began with the mean color-concept association data (Fig. \ref{fig:SuppSunShadeHill}). From the 71 individual colors, there were 2485 possible pairs. To ensure there was obvious lightness variation to activate the dark-is-more bias, we reduced this number of pairs by excluding pairs of the same lightness (637 pairs) or pairs that had a small lightness difference (difference of L* $<38$ in CIELAB space) (812 pairs). It is not yet known how much lightness difference is needed to activate the dark-is-more bias, but this should provide a conservative amount given prior work on color discriminability along the L* axis \cite{szafir2018, stone2014}. We also removed pairs that included white or black (96 pairs) to reduce the possibility that the opaque-is-more bias would be activated \cite{schloss2019} . After this exclusion, we limited the set by only working with pairs that had a lightness difference of either 38 or 50 (143 pairs excluded).

For the remaining color pairs, we linearly interpolated between the light and dark colors in CIELAB space to create a color scale with 10 total steps (Fig. \ref{fig:exp1methods}). Ideally, association strengths between the domain concepts and the interpolated colors would fall between association strengths for endpoint colors. To estimate the direct associations for intermediate colors, we used a color space regression model as in Schloss et al. \cite{schloss2018modeling}. Color space regression models predict a dependent measure (e.g., color preferences in \cite{schloss2018modeling}) from coordinates of each color in a well-specified color space. Schloss et al.  \cite{schloss2018modeling} tested several kinds of color space models and found that a cylindrical model in CIELCh space with two harmonics best predicted color preference data, so we use that model here. The model includes seven predictors for each color: lightness (L*), chroma (C*), first harmonic of hue angle ($\sin(h)$ and $\cos(h)$), second harmonic of hue angle ($\sin(2h)$ and $\cos(2h)$), and an additive constant. We note that CIELCh space is a cylindrical representation of the coordinates in CIELAB space. Lightness (L*) coordinates are the same in both spaces, whereas a* and b* in CIELAB (roughly) represent red/green and blue/yellow Cartesian axes, C* and h in CIELCh represent radius and angle. 

We used the following procedure to predict associations for colors for which we did not have direct association ratings. First, for each concept, we conducted a multiple linear regression model to estimate the mean direct association ratings for the colors we did have (UW-71). This analysis provided weights for the seven predictors and the additive constant for each concept. We checked that the models fit the data well, and found strong fits for both sunshine (multiple $R = .81$) and shade (multiple $R =  .88$).  Second, we used these weights and constant to define the regression equation for each concept. Finally, we plugged the CIE L*, C*, and $h$ coordinates for the new, interpolated colors into the regression equation for each concept to estimate the direct color-concept associations. 

\begin{figure}
 \centering
 \includegraphics{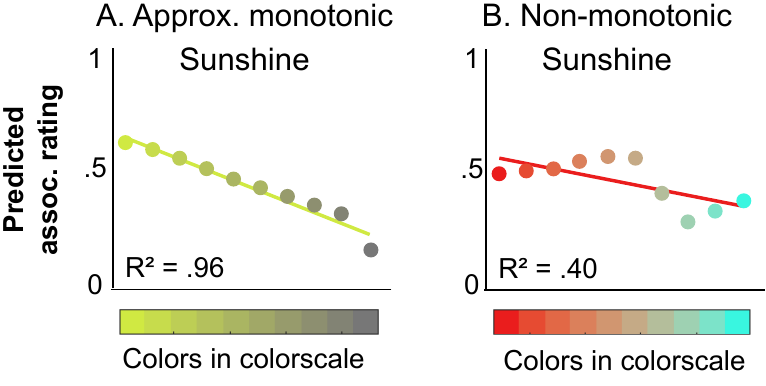}
 \caption{Predicted association ratings (circles filled according to color scale) for colors within two color scales representing amounts of sunshine. In (A) the predicted associations are approximately monotonic, whereas in (B), the predicted associations are non-monotonic). Color scales ordered with more associated endpoint on the left and less associated endpoint on the right.} \label{fig:SuppMono}
 \vspace{-4mm}
\end{figure}

Once we had the estimated direct associations for interpolated colors between the endpoints of each color scale, we used another linear regression analysis to estimate monotonicity. For each color scale (two endpoints plus eight intermediate colors), we fit a regression line estimating predicted direct association ratings from color sequence (1-10) in the color scale (Fig. \ref{fig:SuppMono}). If model estimates for interpolated colors were not intermediate between associations strengths for endpoint colors (non-monotonic), there should be a poor fit (low $R^{2}$). We set the exclusion criterion for color scales to be to be $R^{2} < .8$. We note that this estimate of monotonicity is a heuristic, and it is possible for model estimates to vary monotonically while being poorly fit by a regression line. Thus, our criterion may have been overly conservative in excluding candidate color scales.   

\begin{figure}
 \centering
 \includegraphics{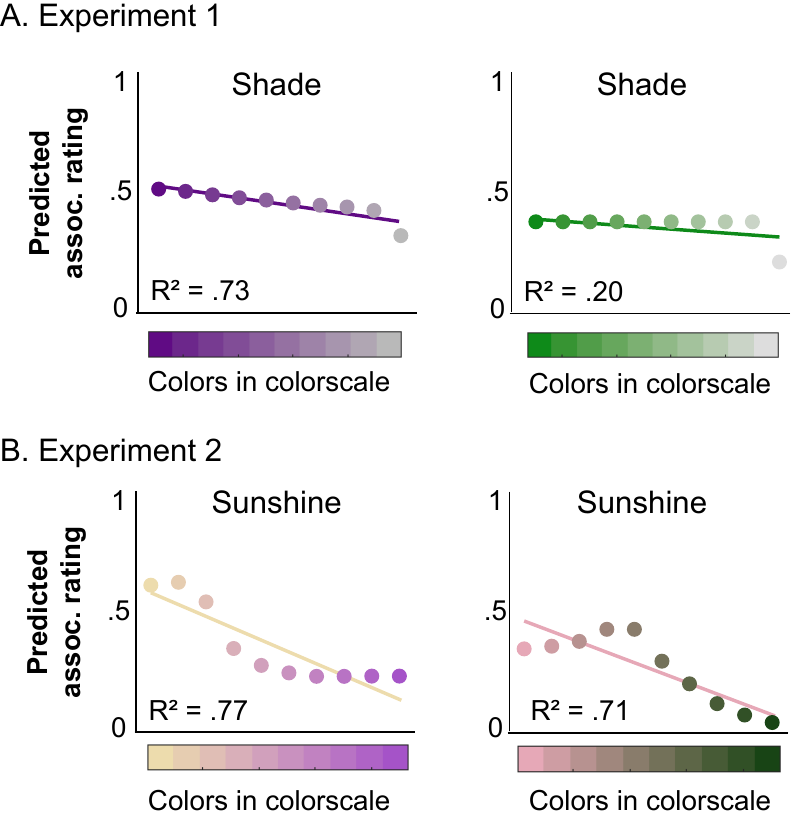}
 \caption{Predicted association ratings (circles filled according to color scale) for colors within color scales that did not pass the monotonicity heuristic criterion. (A) Predicted associations for shade (Exp. 1) do still appear to vary monotonically in association strength. (B) Predicted associations for sunshine (Exp. 2) vary such that intermediate colors are more strongly associated than the endpoints. Color scales ordered with more associated endpoint on the left and less associated endpoint on the right.} \label{fig:SuppMonoShadeSun}
 \vspace{-4mm}
\end{figure}

A total of 429 pairs of endpoint color pairs were excluded for not passing this criterion. Fig. \ref{fig:SuppMono}A shows an example color scale that meets this criterion of approximate monotonicity for the concept sunshine. The estimated associations for the interpolated colors fall between the associations expected for the endpoints colors, meaning the lightest color is predicted to be most associated with sunshine, and the darkest color is predicted as the least associated. The drop-off that happens for the last point (darkest gray) could be due to chroma equaling zero, and consequently dropping out of the equation. This drop-off for an achromatic endpoint emerges for other color scales (see Fig. \ref{fig:SuppMonoShadeSun}A). Fig. \ref{fig:SuppMono}B shows a color scale that is estimated to be non-monotonic for sunshine. Here, some interpolated colors are expected to be more associated with sunshine than either endpoint color of the scale.  

When we originally performed this linear regression analysis to estimate monotonicity, there was an error in the code, in which hue angle was treated as radians, rather than degrees. This resulted in slightly different estimated associations and fits. Once corrected, two color scales for shade did not meet our statistical criterion of a fit of $R^{2} > .8$ in Exp. 1. Our statistical criterion is just a heuristic, and visual inspection suggested that the intermediate colors still vary monotonically such that the interpolated colors were not more or less associated than the endpoints, so we kept them in the analyses (see Fig. \ref{fig:SuppMonoShadeSun}A).

We selected our final eight color scales (Fig. \ref{fig:exp1methods}) by identifying four high association difference color scales and four low association difference color scales. The high association difference color scales were those that had the largest association difference for both sunshine and shade. The low association difference color scales had association differences near zero for both concepts. When making this final selection, we chose color scales that appeared different from one another to provide variability in our stimulus set.

\begin{figure*}
 \centering
 \includegraphics{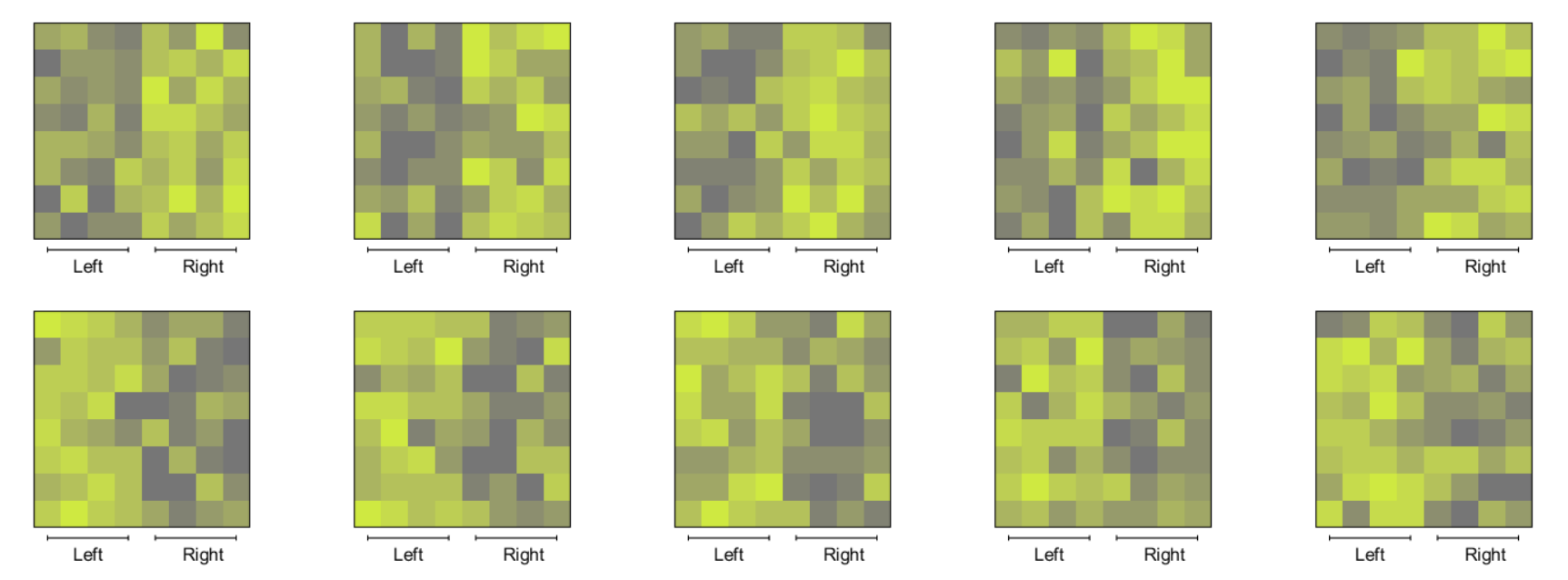}
 \caption{Colormap data visualizations generated using the 10 underlying datasets in study. In the top row, the darker side is on the left, and in the bottom row, the darker side is on the right. The examples in this figure were created using one of the color scales from Exp. 1. Different color scales were applied to these same 10 underlying datasets to create all the colormap stimuli for this study.}
 \label{fig:SuppExp1colormaps}
 \vspace{-4mm}
\end{figure*}

We applied each color scale to 10 underlying datasets to create a set of 80 colormaps. Half the datasets featured the darker color on the left side (as in Fig. \ref{fig:SuppExp1colormaps} top) and the other half featured the darker color on the right side (as in Fig. \ref{fig:SuppExp1colormaps} bottom).

\textbf{Experiment 2.}
We used the same approach as in Exp. 1, with two exceptions. First, the only domain concept in this experiment was sunshine, so we only checked monotonicity for sunshine, and not shade.  

After exclusions, we examined the association differences between the endpoint colors for the remaining color scales and  selected 8 new color scales that spanned the full range of association differences between the high and low color scales used in Exp. 1. We also checked that colormaps generated using these colors scales appeared different from each other. To create the colormaps for this experiment, we applied these new color scales to the same 10 underlying datasets. 

After collecting the data for Exp. 2, we discovered that two of the color scales in this experiment did not meet our statistical criterion for monotonicity due to the same coding error treating hue angle as radians instead of degrees as described for Exp. 1. Visual inspection showed that these color scales strongly deviated from the best fit line, and in one case the intermediate colors were more associated with sunshine than the endpoints (Fig. \ref{fig:SuppMonoShadeSun}B). Thus, we excluded data from these two color scales from our analyses in Exp. 2.

\textbf{Experiment 3.} We used the same approach as in Exp. 1, with the following exceptions. When selecting color pairs to generate the color scales, we allowed black and white to be included to provide more possible color pairs to choose from. Fig  \ref{fig:SuppEnvironhill} shows the mean direct association ratings for each of the environmental concepts. 
Unique to this experiment, we investigated the full range of signed association difference within each domain concept. This full range of signed association difference corresponded to a full range of signed semantic distance (when non-zero weight was placed on merit for direct associations). To select color scales spanning this range, we grouped all remaining color scales into three even-sized bins based on their signed association difference. We selected seven color pairs from each bin, for twenty-one color scales per domain concept that varied along the range of association differences, such that for half the color scales, the direct associations and dark-is-more bias were consistent (Fig. \ref{fig:exp3colorscales}; right color scales) and for half, the direct associations and dark-is-more bias were conflicting (Fig. \ref{fig:exp3colorscales}; left color scales). 

During this process, we narrowed our set of concepts to glacial ice, ocean water, and wild fire given there were color pairs for these concepts that spanned nearly the full range of signed association differences, whereas the color pairs for tree foliage and ground soil were more limited.  We the checked that the color space regression model strongly fit our chosen domain concepts (multiple $R = .94$ for wild fire, multiple $R = .90$ for ocean water, and multiple $R = .90$ for glacial ice). We also ensured that interpolations between each pair of selected end point colors passed the monotonicity criterion.

\subsection{Additional data for estimating merit}\label{sec:SuppAddData}
For each domain concept used in the colormap tasks, we collected direct association ratings for the endpoint quantities of each domain concept (``no [domain concept]'' and ``a lot of [domain concept]''), which was displayed at the top of the screen. We used the means of these ratings to specify merit for direct associations. For Exp. 3, we also collected darkness difference ratings which we used to specify merit for the dark-is-more bias. 

\textbf{Sunshine endpoint associations.}
This task was the same as the direct associations task described in Section \ref{sec:Exp1assoc}, with the following exceptions.  Instead of rating associations for all UW-71 colors with the domain concept name  (e.g., sunshine), participants only rated their associations for each endpoint color in the color scales and each endpoint quantity of the domain concept (e.g., ``no sunshine'' and ``a lot of sunshine''). Participants judged each of the endpoint colors from the color scales in Exp. 2, which included endpoint colors from Exp. 1 as well.  

Before beginning the task, participants were shown the two endpoint concepts and the set of endpoint colors, presented as individual squares. To anchor the endpoints of the response scale, they were prompted to think about which color they associated most and least for each endpoint concept (``no sunshine'' and ``a lot of sunshine''). 

 After completing the block for sunshine, participants completed the attention check as used in Exp. 3. For each endpoint concept, we transformed the ratings from a scale of -200 to 200, to a scale of 0 to 1. We then averaged the mean across all participants. 

The target sample size was $n=30$ as in Exp. 1. We collected 30 participants initially, determined how many participants were excluded, and iteratively posted the experiment to collect data for the number of participants excluded until reaching at least 30 participants. Exclusions were based on atypical color vision (as assessed in Exp. 2) and the attention check (as used in Exp. 3) (48 Amazon mTurk workers in total, 17 excluded for atypical color vision, 0 for failing the attention check).  Our final sample included 31 participants (mean age = 39 years old; 20 women, 11 men). The experiment took approximately 10 min. and participants were compensated with \$1.21.

\textbf{Shade endpoint associations.}
We also collected endpoint associations for shade, but they were not used in any of analyses in this paper. This task was the same as the endpoint association ratings task for sunshine, except ``sunshine'' was replaced with ``shade'', and it only included the endpoint colors from Exp. 1.  

The target sample size was $n=30$. We collected 30 participants initially, determined how many participants were excluded, and iteratively posted the experiment to collect data for the number of participants excluded until reaching at least 30 participants. Exclusions were based on atypical color vision and the attention check (56 Amazon mTurk workers in total, 10 excluded for atypical color vision, 16 for failing the attention check). Our final sample size was 30 participants (mean age = 39 years old; 10 women, 20 men). The average time to complete the experiment was approximately 5 min. Participants were compensated with \$0.60.

\textbf{Environment concept endpoints.} We collected endpoint direct association ratings for the three domain concepts selected for Exp. 3: wild fire, glacial ice, and ocean water. Domain concepts varied within-subject, presented in a blocked randomized design, such that participants rated their associations for the two endpoint concepts for a given domain concept, prior to moving on to the next domain concept. The two endpoint concepts were presented in a random order within each domain concept. At the start of each block, participants were shown the endpoint concept names and the set of endpoint colors they would be judging in that block to anchor the endpoints of the response scale.

The target sample size was $n=30$. We collected 30 participants initially, determined how many participants were excluded, and iteratively posted the experiment to collect data for the number of participants excluded until reaching at least 30 participants. Exclusions were based on atypical color vision and the attention check (37 undergraduate psychology students in total, 5 excluded for atypical color vision, 5 for failing the attention check). Our final sample size was 30 participants (mean age = 19 years old; 25 women, 4 men, 1 non-binary). The experiment took approximately 30 min. and participants received course credit.

 \textbf{Darkness difference ratings}
As described in the Section \ref{sec:exp3_maps_merit}) of the main text, we collected darkness difference ratings to quantify the degree to which one endpoint color was clearly darker than the other endpoint color within each color scale. The means of these ratings were used to quantify merit for the dark-is-more bias (see Section \ref{sec:exp3_maps_merit}).

We collected data from four volunteers with expertise in color perception. During the task, participants saw two colored squares above the sliding scale with the endpoints of the scale labeled ``left color is clearly darker'' and ``right color is clearly darker''. The middle point of the scale was labeled ``equal darkness.'' Participants used the slider scale to indicate which color in the pair they thought was darker, and the degree to which they thought that color was darker. Participants rated each of the color pairs twice, once with the darker color on the left and once with the darker color on the right. 

At the beginning of the experiment, participants were shown an achromatic color gradient (5 steps from black to white) and were instructed how to interpret darkness. To indicate what we meant by darkness, participants were told that the color gradient displayed ranged from black to white. Colors near the left were darker, with black being the darkest, and colors near the right were lighter, with white being the lightest. They were then told that all colors fall somewhere along this dimension ranging for dark to light and to keep this idea of darkness in mind during the experiment. We averaged ratings across the two repetitions for each color pair for each participant, and then averaged across participants. The experiment took approximately 15 min. and participants were not compensated.

\subsection{Power analyses for colormap tasks}\label{sec:SuppPower}
To conduct power analyses for our logistic mixed-effect regression model, we used the mixed power package in R \cite{kumle2021}.

\textbf{Experiment 1.} We conducted the power analysis using a logistic mixed-effects regression model with data from a different study. The model predicted which side of an achromatic colormap participants selected as representing more alien animal sightings (left = 1, right = 0) from a factor coding which side was darker (left darker = 1, right darker = -1). To achieve a power of at least .8, we aimed to recruit 12 participants per condition (total $n=192$).

\textbf{Experiment 2}
We conducted the power analysis using data from participants in Exp. 1 who judged low association difference colormaps representing sunshine. The analysis was based on a binomial logistic mixed-effect regression predicting the side selected (left = 1, right = 0) from which side was dark (left darker = 1, right darker = -1), and a by-subject random intercept. The goal was to have enough participants per condition to support stable differences from chance, should they exist, while examining a fuller range of association differences. To achieve a power of at least .8, recruited 42 participants per condition, which would also account for expected exclusions for atypical color vision. Our target sample size was $n=640$. 

\textbf{Experiment 3.} We conducted a power analysis with the data from Exp. 2, using a binomial logistic regression in which we predicted which side was selected (left = 1, right = 0) from which side was dark (left darker = 1, right darker = -1) and the signed semantic distance for the pair (ranging from light-more associated: -1 to dark-more associated: 1). To achieve a power of at least .8, we aimed to recruit 20 participants per condition, which would also support having enough participants to split the data into a testing set and training set. The target sample size was $n=1260$.

\subsection{Additional analyses}\label{sec:SuppAddAnalysis}
\subsubsection{Experiment 1}
An alternative analysis for evaluating the results from Exp. 1 is to use the domain concept as a factor, and code the association difference as positive, regardless of which color is more associated. In doing so, we can still test for independent effects of direct associations and the dark-is-more bias. 

With these predictors, we used a logistic mixed-effects regression to predict participant's response (left vs. right) from the concept (shade vs. sunshine), the average (unsigned) association difference (centered), and which side was dark (dark-left vs. dark-right). Here, we include all pair-wise interactions, and the three way interaction, as well a by-subject random intercept (Table \ref{tab:expFull}). Overall, participants chose the darker side, consistent with the dark-is-more bias. The significant three-way interaction suggests responses varied depending on the concept and color pairs used to create the colormap.

  \begin{table}[ht!!]
  \caption{Logistic mixed-effects model predicting side chosen from dark side (dark), association difference (assoc), concept, and all interactions.}
  \label{tab:expFull}
  \begin{tabular}{lrrrr}  
  \toprule
 \bf{Factor} & $\pmb{\beta}$ & \bf{SE} & $\pmb{z}$ & $\pmb{p}$ 
  \\
  \midrule
   Intercept & $.065$ & $.103$ & $.625$ & $.532$
  \\ 
  Dark & $ 1.785$ & $.207$ & $8.639$ & $<.001$
  \\ 
   Assoc & $.293$ & $.330$ & $.889$ & $.374$
  \\ 
  Concept & $ -.170$ & $.207$ & $-.823$ &$.410$
  \\
  Dark*Assoc & $-4.821$ & $.660$ & $-7.303$ & $<.001$
  \\ 
   Dark*Concept & $9.164$ & $.413$ & $22.18$ & $<.001$
  \\ 
   Assoc*Concept & $-.201$ & $.660$ & $-.305$ & $.760$
  \\ 
 Dark*Assoc*Concept & $11.67$ & $1.321$ & $8.833$ &$<.001$
 \\
 \bottomrule
  \end{tabular}
\end{table}

 We ran the model separately for shade and sunshine to better understand this three-way interaction, and consequently the interactions of dark side $\times$ concept and dark side $\times$ association difference (Table \ref{tab:expConcept}). For shade, dark side was the only significant predictor. This indicates the darker side was more likely to be selected as having more shade, regardless of direct associations of the colors in the colormap.
 
For the concept sunshine, on average, participants also selected the darker side more often. However, there was a significant interaction of dark side and association difference. This interaction indicates that for low association difference maps, participants were more likely to select the darker side as having more sunshine, even though sunshine is generally associated with lighter colors. However, for high association difference maps, participants were more likely to select the lighter side as having more sunshine. This pattern of results aligns with those discussed in Section \ref{sec:Exp1mapResults} and indicates that direct color-concept associations can override the dark-is-more bias, and most closely resembles the prediction that both the dark-is-more bias and direct associations influence people's inferred mappings (Fig. \ref{fig:exp1results}A).

 \begin{table}[h!]
  \caption{Logistic mixed-effects models predicting side chosen from dark side (dark), association difference (assoc), and their interaction for each concept.}
  \label{tab:expConcept}
  \begin{tabular}{llrrrr}  
  \toprule
 \bf{Concept} & \bf{Factor} & $\pmb{\beta}$ & \bf{SE} & $\pmb{z}$ & $\pmb{p}$ 
  \\
  \midrule
  Shade &  Intercept & $-.021$ & $.170$ & $-.121$ & $.904$ \\ 
        & Dark & $6.366$ & $.339$ & $18.772$ & $<.001$\\ 
        & Assoc & $.193$ & $.542$ & $.356$ & $.772$\\
        & Dark*Assoc & $1.012$ & $1.084$ & $.934$ & $.350$\\
  \midrule
  Sunshine  &  Intercept & $.150$ & $.118$ & $1.268$ & $.205$\\
            & Dark & $-2.797$ & $.236$ & $-11.86$ & $<.001$\\ 
            & Assoc & $.394$ & $.377$ & $1.05$ & $.296$\\ 
            & Dark*Assoc & $-10.66$ & $.753$ & $-14.15$ & $<.001$\\ 
  \bottomrule
  \end{tabular}
  \vspace{-4mm}
\end{table}

\subsubsection{Experiment 3} \label{sec:Exp3SupResults}
In Exp. 3 we aimed to develop and test a method for combining multiple sources of merit. Here, we include additional details of the analysis reported in the main text, and an additional analysis, comparable to those performed for Exp. 1 and 2, that allow for testing the independent effects of direct associations and the dark-is-more bias. 

As described in Section \ref{sec:Exp3Results} of the main text, we used the data from the testing set to evaluate whether the optimal weight pair was better for predicting assignment inference than each source of merit alone. To examine effects of relative weighting, we used a linear mixed effects model predicting MSE for each color scale, with fixed effects for relative weighting, domain concept, and their interaction (using Helmert contrasts). The model also included a by-color scale random intercept and random slope for relative weighting. 

There was a main effect of relative weighting ($F(2,53.45) = 9.18, p <.001$), which was due to lower MSE for combined merit than for the average of each source of merit alone ($t(53.55) = -4.24, p <.001$). Performance of each source of merit alone did not significantly differ ($t(48.32) = -1.14, p = .26$). Our model also revealed a main effect of concept ($F(2,60.76) = 12.66, p <.001$). Fits were significantly better for wild fire compared to the average of water and ice ($t(91.75) = -5.01, p <.001$), with no significant difference between ocean water and glacial ice ($t(100.08) = .68, p = .50$). Concept interacted with relative weighting ($F(4,65.63) = 3.13, p = .02$), which was driven by the MSE for fire being especially low for the optimal weight pairing compared to the average of the other pairings ($t(66.84) = 2.57, p = .01$), and being lower for direct associations alone compared to dark-is-more alone ($t(103.15) = 2.87, p = .005$).

To test for independent effects of direct associations and the dark-is-more bias, we also used a binomial logistic regression to predict the side selected (left vs. right) from a predictor coding which side was more associated and the degree to which it was more associated measured by semantic distance using association difference as merit (ranged from -1 to 1), and a predictor coding which side was darker and the degree it was darker determined by darkness difference ratings (scaled to range from -1 to 1), and a by-subject random intercept and random effects for each fixed factor. Signed association semantic distance ($B = 16.91, SE = 1.35; z = 12.56, p < .001$) and darkness ratings ($B = 13.66 SE = 1.25; z = 10.91, p < .001$) both emerged as significant predictors, further supporting that both contribute independent effects on inferred mappings.

\begin{figure} [tb]
 \centering
 \includegraphics{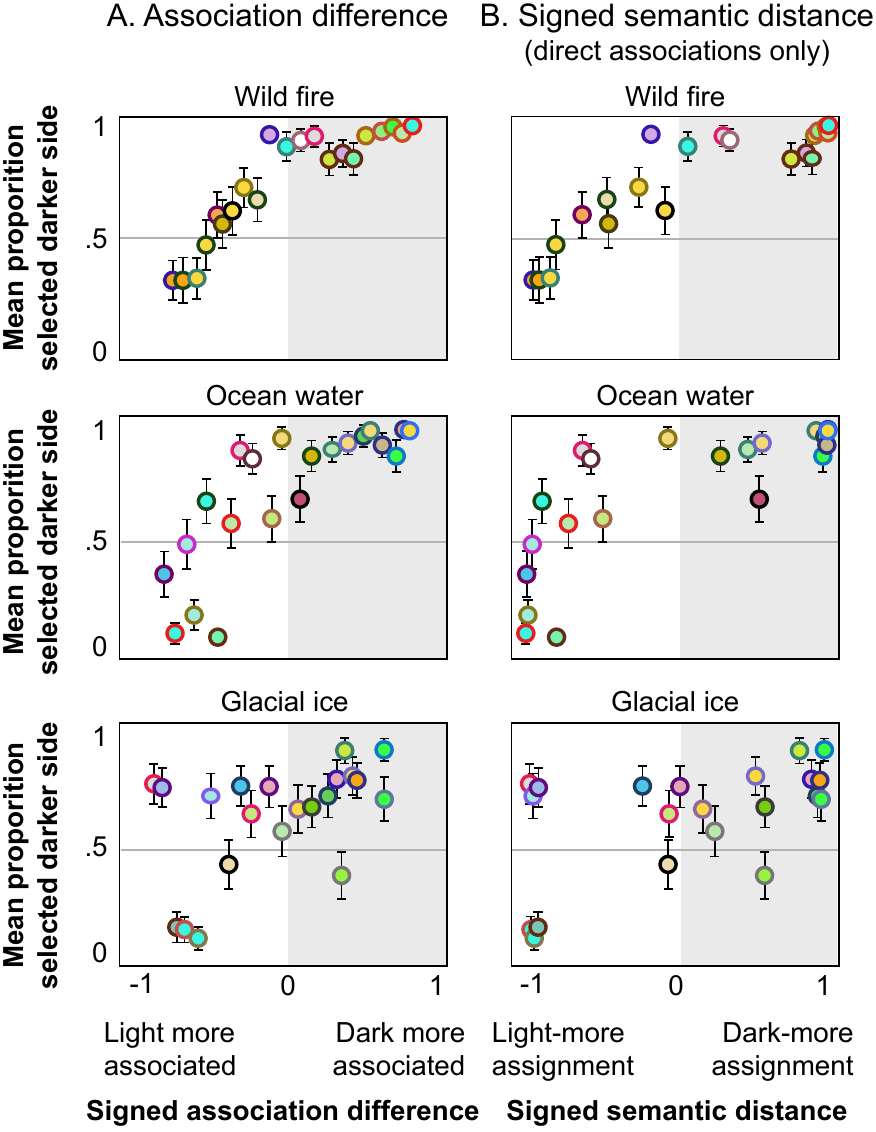}
 \caption{Mean proportion of times the darker side was selected, plotted as a function of (A) signed association difference, and (B) signed semantic distance using direct associations as merit, for each domain concept: wild fire, ocean water, and glacial ice. Each mark represents a color scale, and the stroke and fill colors represent the endpoint colors of the color scales. Error bars represent standard errors of the means.}
 \label{fig:SuppExp3assocDist}
 \vspace{-5mm}
\end{figure}

Recall that in Exp. 2, once association difference reached about $-.55$, participants almost always inferred that the lighter side of the colormaps mapped to more of the domain concept (sunshine). This plateau could be explained by semantic distance reaching its limit (Fig. \ref{fig:exp2results}). Fig. \ref{fig:SuppExp3assocDist} shows analogous plots for Exp. 3. It shows the mean proportion of times participants selected the darker side for colormaps generated from each color scale for each domain concept, as function of (A) association difference and (B) signed semantic distance based on direct associations. The data for these concepts do not appear to have the plateau around $-.55$. 

We propose that the reason for this difference is that sunshine is more inherently associated with light colors than the concepts in  Exp. 3. That is, the correlation between lightness (CIELAB L*) of the 71 colors and mean association ratings for sunshine was strong ($r=.71$), whereas those correlations are weak for wild fire ($r=-.03$), ocean water ($r=.05$), and glacial ice ($r=.33$). Direct associations may be less effective at overriding the dark-is-more bias for concepts that are less inherently light. If so, then it may be necessary to build inherent lightness into simulations of assignment inference in the future. 

We also note that Exp. 3 included fewer color scales in the association difference range $-.55$ to $-1$ where the plateau occurred in Exp. 2. As discussed in the main text, glacial ice has extreme outliers in which the dark-is-more bias strongly appears to override direct associations when they conflict. These outliers may be due to activation of the opaque-is-more bias, but future work designed to test the opaque-is-more bias will be necessary to study its effects.

\begin{table*}[!hb]
\sisetup{round-mode=places}
\centering
\caption{Coordinates for the University of Wisconsin 71 (UW-71) colors in CIE 1931 xyY, CIELAB, and CIELCh space. The white point used to convert between CIE 1931 xyY and CIELAB space was CIE Illuminant D65 (x = 0.313, y = 0.329, Y = 100). Table adapted from \cite{mukherjee2022}.}
\label{table:UW_71_colors}
\renewcommand{\arraystretch}{0.9}
\begin{tabular}
{c*{1}{
    S[round-precision=3]
    S[round-precision=3]
    S[round-precision=2]
    S[round-precision=2]
    S[round-precision=2]
    S[round-precision=2]
    S[round-precision=2]
    S[round-precision=2]
}}
\toprule
\multicolumn{1}{c}{\textbf{Color}} &
\multicolumn{1}{c}{\textbf{x}} & \multicolumn{1}{c}{\textbf{y}} & \multicolumn{1}{c}{\textbf{Y}} & \multicolumn{1}{c}{\textbf{L*}} & \multicolumn{1}{c}{\textbf{a*}} & \multicolumn{1}{c}{\textbf{b*}} &
\multicolumn{1}{c}{\textbf{C*}} & \multicolumn{1}{c}{\textbf{h}} \\
\midrule
1 & 0.17813 & 0.14021 & 18.419 & 50 & 28.891 & -73.589 & 79.057 &	291.435\\
2 & 0.1742 & 0.082514 & 4.4155 & 25 & 53.857 & -72.28 & 90.139 & 306.690\\
3 & 0.21726 & 0.13588 & 18.419 & 50 & 53.857 & -72.28 & 90.138 & 306.690\\
4 & 0.2591 & 0.13088 & 18.419 & 50 & 78.822 & -70.972 & 106.066 & 318.000\\
5 & 0.18715 & 0.19157 & 18.419 & 50 & 2.6168 & -49.931 & 50.000 & 273.000\\
6 & 0.19063 & 0.1298 & 4.4155 & 25 & 27.583 & -48.623 & 55.902 & 299.566\\
7 & 0.23145 & 0.18448 & 8.419 & 50 & 27.583 & -48.623 & 55.902 & 299.566\\
8 & 0.25495 & 0.12284 & 4.4155 & 25 & 52.548 & -47.315 & 70.711 & 318.000\\
9 & 0.27872 & 0.17635 & 18.419 & 50 & 52.548 & -47.315 & 70.711 & 318.000\\
10 & 0.32783 & 0.1674 & 18.419 & 50 & 77.514 & -46.006 & 90.139 & 329.310\\
11 & 0.22397 & 0.28399 & 48.278 & 75 & -23.657 & -26.274 & 35.355 &	228.000\\
12 & 0.2081 & 0.21415 & 4.4155 & 25 & 1.3084 & -24.966 & 25.000 & 273.000\\
13 & 0.24471 & 0.25395 & 18.419 & 50 & 1.3084 & -24.966 & 25.000 & 273.000\\
14  & 0.26261 & 0.27354 & 48.278 & 75 & 1.3084 & -24.966 & 25.000 & 273.000\\
15  & 0.28644 & 0.19884  & 4.4155 & 25 & 26.274 & -23.657 & 35.355 &	318.000\\
16 & 0.29797 & 0.24051  & 18.419 & 50 & 26.274 & -23.657  & 35.355 & 318.000\\
17 & 0.30272 & 0.2622  & 48.278  & 75 & 26.274 & -23.657  & 35.355 & 318.000\\
18 & 0.36941 & 0.18108  & 4.4155 & 25  & 51.24 & -22.349 & 55.902 &	336.435\\
19 & 0.35288 & 0.2259 & 18.419 & 50 & 51.24 & -22.349 & 55.902 & 336.435\\
20  & 0.40795 & 0.21059 & 18.419 & 50 & 76.206 & -21.041 & 79.057 & 344.565\\
21 & 0.23784 & 0.35662 & 72.065 & 88 & -49.931 & -2.6168 & 50.000 &	183.000\\
22 & 0.25332 & 0.35108 & 18.419 & 50 & -24.966 & -1.3084 & 25.000 & 183.000\\
23 & 0.26938 & 0.34523 & 48.278 & 75 & -24.966 & -1.3084 & 25.000 & 183.000\\
24 & 0.27473 & 0.34327 & 72.065 & 88 & -24.966 & -1.3084 & 25.000 & 183.000\\
25 & 0.313 & 0.3290 & 0 & 0 & 0 & 0 & 0 & 0\\
26 & 0.31273 & 0.32902 & 4.4155 & 25 & 0 & 0 & 0 & 0\\
27 & 0.31273 & 0.32902 & 18.419 & 50 & 0 & 0 & 0 & 0 \\
28 & 0.31273 & 0.32902 & 48.278 & 75 & 0 & 0 & 0 & 0 \\
29 & 0.31273 & 0.32902 & 100.00 &  100 & 0 & 0  & 0 & 0\\
30 & 0.31273 & 0.32902 & 72.065 & 88 & 0 & 0 & 0 & 0 \\
31 & 0.41044 & 0.2905 & 4.4155 & 25 & 24.966 & 1.3084 & 25.000 &	3.000\\
32 & 0.37353 & 0.30534 & 18.419 & 50 & 24.966 & 1.3084 & 25.000 & 3.000\\
33 & 0.3568 & 0.31196 & 48.278 & 75 & 24.966 & 1.3084 & 25.000 &	3.000 \\
34 & 0.43376 & 0.28095 & 18.419 & 50 & 49.931 & 2.6168 & 50.000 & 3.000 \\
35 & 0.49181 & 0.25666  & 18.419 & 50 & 74.897 & 3.9252 & 75.000 & 3.000\\ 
36 & 0.27022 & 0.43268 & 48.278 & 75 & -51.24 & 22.349 & 55.902 & 156.435\\
37 & 0.27623 & 0.41829 & 72.065 & 88 & -51.24 & 22.349 & 55.902 & 156.435\\
38 & 0.3075 & 0.52435 & 4.4155 & 25  & -26.274 & 23.657 & 35.355 & 138.000\\
39 & 0.31561 & 0.44408 & 18.419 & 50  &  -26.274 & 23.657 & 35.355 & 138.000\\
40 & 0.31654 & 0.41004 & 48.278 & 75 & -26.274 & 23.657 & 35.355 & 138.000\\
41 & 0.31652 & 0.39917 & 72.065 & 88 & -26.274 & 23.657 & 35.355 & 138.000\\
42 & 0.41791 & 0.4496 & 4.4155 & 25 & -1.3084 & 24.966 & 25.000 & 93.000\\
43 & 0.38179 & 0.40728 & 18.419 & 50 & -1.3084 & 24.966 & 25.000 & 93.000\\
44 & 0.36355 & 0.38634 & 48.278 & 75 & -1.3084 & 24.966 & 25.000 & 93.000\\
45 & 0.35735 & 0.37928 & 72.065 & 88 & -1.3084 & 24.966 & 25.000 & 93.000\\
46 & 0.52174 & 0.37656 & 4.4155 & 25 & 23.657 & 26.274 & 35.355 & 48.000\\
47 & 0.44682 & 0.36993 & 18.419 & 50 & 23.657 & 26.274 & 35.355 & 48.000\\
48 & 0.41032 & 0.36214 & 48.278 & 75 & 23.657 & 26.274 & 35.355 & 48.000\\
49 & 0.50873 & 0.33341 & 18.419 & 50 & 48.623 & 27.583 & 55.902 & 29.566\\
50 & 0.56618 & 0.29873 & 18.419 & 50 & 73.589 & 28.891 & 79.057 & 21.435\\
51 & 0.29736 & 0.57731 & 18.419 & 50 & -52.548 & 47.315 & 70.711 & 138.000\\
52 & 0.31049 & 0.50294 & 48.278 & 75 & -52.548 & 47.315 & 70.711 & 138.000\\
53 & 0.31313 & 0.47888 & 72.065 & 88 & -52.548 & 47.315 & 70.711 & 138.000\\
54 & 0.36753 & 0.52508 & 18.419 & 50 & -27.583 & 48.623 & 55.902 & 119.566\\
55 & 0.35998 & 0.47135 & 48.278 & 75 & -27.583 & 48.623 & 55.902 & 119.566\\
56 & 0.35593 & 0.45307 & 72.065 & 88 & -27.583 & 48.623 & 55.902 & 119.566\\
57 & 0.43671 & 0.47238 & 18.419 & 50 & -2.6168 & 49.931 & 50.000 & 93.000\\
58 & 0.4092 & 0.43925 & 48.278 & 75 & -2.6168 & 49.931 & 50.000 & 93.000\\
59 & 0.39861 & 0.42682 & 72.065 & 88 & -2.6168 & 49.931 & 50.000 & 93.000\\
60 & 0.50246 & 0.42134 & 18.419 & 50 & 22.349 & 51.24 & 55.902 & 66.435\\
61 & 0.4572 & 0.40735 & 48.278 & 75 & 22.349 & 51.24 & 55.902 & 66.435\\
62 & 0.56315 & 0.37345 & 18.419 & 50 & 47.315 & 52.548 & 70.711 & 48.000\\
63 & 0.61793 & 0.32963 & 18.419 & 50 & 72.28 & 53.857 & 90.139 &	36.690\\
64 & 0.30023 & 0.56426 & 72.065 & 88 & -78.822 & 70.972 & 106.066 & 138.000\\
65 & 0.34264 & 0.56147 & 48.278 & 75 & -53.857 & 72.28 & 90.139 & 126.690\\
66 & 0.34455 & 0.53221 & 72.065 & 88 & -53.857 & 72.28 & 90.139 & 126.690\\
67 & 0.39381 & 0.52123 & 48.278 & 75 & -28.891 & 73.589 & 79.057 & 111.435\\
68 & 0.38886 & 0.49964 & 72.065 & 88 & -28.891 & 73.589 & 79.057 & 111.435\\
69 & 0.44388 & 0.48131 & 48.278 & 75 & -3.9252 & 74.897 & 75.000 & 93.000\\
70 & 0.43244 & 0.46714 & 72.065 & 88 & -3.9252 & 74.897 & 75.000 & 93.000\\
71 & 0.49196 & 0.4425 & 48.278 & 75 & 21.041 & 76.206 & 79.057 & 74.565\\
\bottomrule
\end{tabular}
\end{table*}

\end{document}